\documentclass[twocolumn,superscriptaddress,amsmath,amssymb,aps,longbibliography,prx]{revtex4-2}
\usepackage[colorlinks=true,urlcolor=blue,citecolor=blue,linkcolor=blue]{hyperref}
\usepackage{txfonts}
\usepackage{graphicx}
\usepackage{dsfont}
\usepackage{bbm}
\usepackage{booktabs}
\usepackage{amsmath}
\usepackage{algorithm}
\usepackage{algorithmicx}
\usepackage{algpseudocode} 
\usepackage{graphicx}
\usepackage{dcolumn}
\usepackage{rotating}
\usepackage{longtable, booktabs, threeparttablex}

\usepackage{multirow}
\usepackage{threeparttable}
\usepackage{color}
\usepackage{xcolor} 
\usepackage{xr}
\definecolor{lightgreen}{RGB}{144,238,144} 
\definecolor{lightgray}{gray}{0.75}

\newcommand{\Eq}[1]{Eq.~(\ref{#1})}

\renewcommand{\thetable}{\arabic{table}}

\begin{document}

\title{Space Group Informed Transformer for Crystalline Materials Generation}

\author{Zhendong Cao}
\affiliation{Beijing National Laboratory for Condensed Matter Physics and Institute of Physics, \\Chinese Academy of Sciences, Beijing 100190, China}
\affiliation{School of Physical Sciences, University of Chinese Academy of Sciences, Beijing 100190, China}

\author{Xiaoshan Luo}
\affiliation{Key Laboratory of Material Simulation Methods and Software of Ministry of Education, College of Physics, Jilin University, Changchun 130012, China}
\affiliation{State Key Laboratory of Superhard Materials, College of Physics, Jilin University, Changchun 130012, P. R. China}


\author{Jian Lv}
\email{lvjian@jlu.edu.cn}
\affiliation{Key Laboratory of Material Simulation Methods and Software of Ministry of Education, College of Physics, Jilin University, Changchun 130012, China}

\author{Lei Wang}
\email{wanglei@iphy.ac.cn}
\affiliation{Beijing National Laboratory for Condensed Matter Physics and Institute of Physics, \\Chinese Academy of Sciences, Beijing 100190, China}
\affiliation{Songshan Lake Materials Laboratory, Dongguan, Guangdong 523808, China}

\date{\today}

\begin{abstract}
We introduce \texttt{CrystalFormer}, a transformer-based autoregressive model specifically designed for space group-controlled generation of crystalline materials. By explicitly incorporating space group symmetry, \texttt{CrystalFormer} greatly reduces the effective complexity of crystal space, which is essential for data-and compute-efficient generative modeling of crystalline materials. Leveraging the prominent discrete and sequential nature of the Wyckoff positions, \texttt{CrystalFormer} learns to generate crystals by directly predicting the species and coordinates of symmetry-inequivalent atoms in the unit cell. We demonstrate the advantages of \texttt{CrystalFormer} in standard tasks such as symmetric structure initialization and element substitution over widely used conventional approaches. Furthermore, we showcase its plug-and-play application to property-guided materials design, highlighting its flexibility. 
Our analysis reveals that \texttt{CrystalFormer} ingests sensible solid-state chemistry knowledge and heuristics by compressing the material dataset, thus enabling systematic exploration of crystalline materials space. The simplicity, generality, and adaptability of \texttt{CrystalFormer} position it as a promising architecture to be the foundational model of the entire crystalline materials space, heralding a new era in materials discovery and design. 
\\
(Received 10 February 2025; Revised 5 May 2025; Accepted 12 September 2025)
\end{abstract}

\maketitle



\section{Introduction}

Machine learning methods are playing an increasingly important role in material discovery, complementing conventional computational approaches~\cite{woodley2008crystal,oganov2019structure}. Generative machine learning, in particular, has been a promising step for matter inverse design~\cite{gomez2018automatic, sanchez2018inverse} which goes beyond machine learning accelerated structure search~\cite{merchant2023scaling} and property screening~\cite{chen2024accelerating}. Generative models learn the underlying distribution of training data and generate new samples from the learned distribution. In addition, the generation process can also be controlled by conditions such as desired material properties or experiment observations. Amazing programming abilities of generative models have been demonstrated in large language model~\cite{achiam2023gpt}, text-to-image generation~\cite{ramesh2021zero, rombach2022high}, and protein design~\cite{ingraham2023illuminating}. 

It is anticipated that generative model-based approaches will introduce groundbreaking changes to the traditional workflows of material discovery. A generative pre-trained foundation model for crystalline materials is a key step towards such a lofty goal. However, despite intensive efforts~\cite{xie2021crystal, luo2023towards, jiao2023crystal, zheng2023predicting, yang2023scalable, zeni2023mattergen, xiao2023invertible, flam2023language, antunes2023crystal, gruver2024finetuned,luo2024deep,ye2024concdvae}, the current generative models for crystalline materials fall short to match the success of other domains. Simply scaling the compute and model size of the current crystal generative model may not be feasible because the amount of high-quality data for crystalline materials is much less compared to language and image domains. Therefore, leveraging the inherent inductive biases specific to crystalline structures for more data-efficient generative modeling is essential, as has been pursued in some of recent works~\cite{zhu2023wycryst, ai4science2023crystal, nguyen2023hierarchical, jiao2024space}.


\begin{table}[t!]
\centering
\renewcommand{\arraystretch}{1.5}
\begin{tabular}{cc}
\toprule
 \textbf{$P1$ world}  & \textbf{With space group symmetry}  \\
\midrule
 $(100\times 100^3)^{20}\approx  10^{160} $ &  $ (100\times 10 \times 100) ^{5  } \approx 10^{25} $   \\
\bottomrule
\end{tabular}
\caption{A back-of-envelope estimate of the size of the crystalline material space. In the ``$P1$ world'', one treats crystals as if they were in the first and the least symmetric $P1$ space group. For the estimate, we consider 100 possible chemical elements and 20 atoms in the unit cell with a coordinate grid size of 100 in each direction. In the case of utilizing the symmetry of a typical space group, we consider 5 symmetry inequivalent atoms occupying 10 possible Wyckoff positions. The additional factor of 100 accounts for the remaining degree of freedom for the fractional coordinates and lattice parameters. See 
Refs.~\cite{oganov2006crystal,davies2016computational} for alternate estimates of the materials space in the context of crystal structure prediction. 
}
\label{tab:count}
\end{table}

The space group symmetry due to the joint outcome of the rotational and translational symmetry in space is arguably the most important inductive bias in the modeling of crystalline materials. There are in total 230 space groups~\cite{glazer2012space} for three-dimensional crystal structures. Nature exhibits a preference for symmetric crystal structures, a tendency that may be attributed to the symmetry inherent in the interatomic interactions, which, in turn, are governed by the fundamental forces acting between elementary particles. As a result, the appearance of crystalline materials in the first and the least symmetric space group $P1$ is rare~\cite{urusov2009frequency}, with many instances potentially even being misclassified~\cite{marsh1999p1}. Failing to match the space group distribution of nature in machine learning-generated materials is regarded as a matter of serious concern~\cite{cheetham2024artificial}.

\begin{figure*}
    \centering
    \includegraphics[trim={5cm 0cm 5cm 0cm}, clip, width=\linewidth]{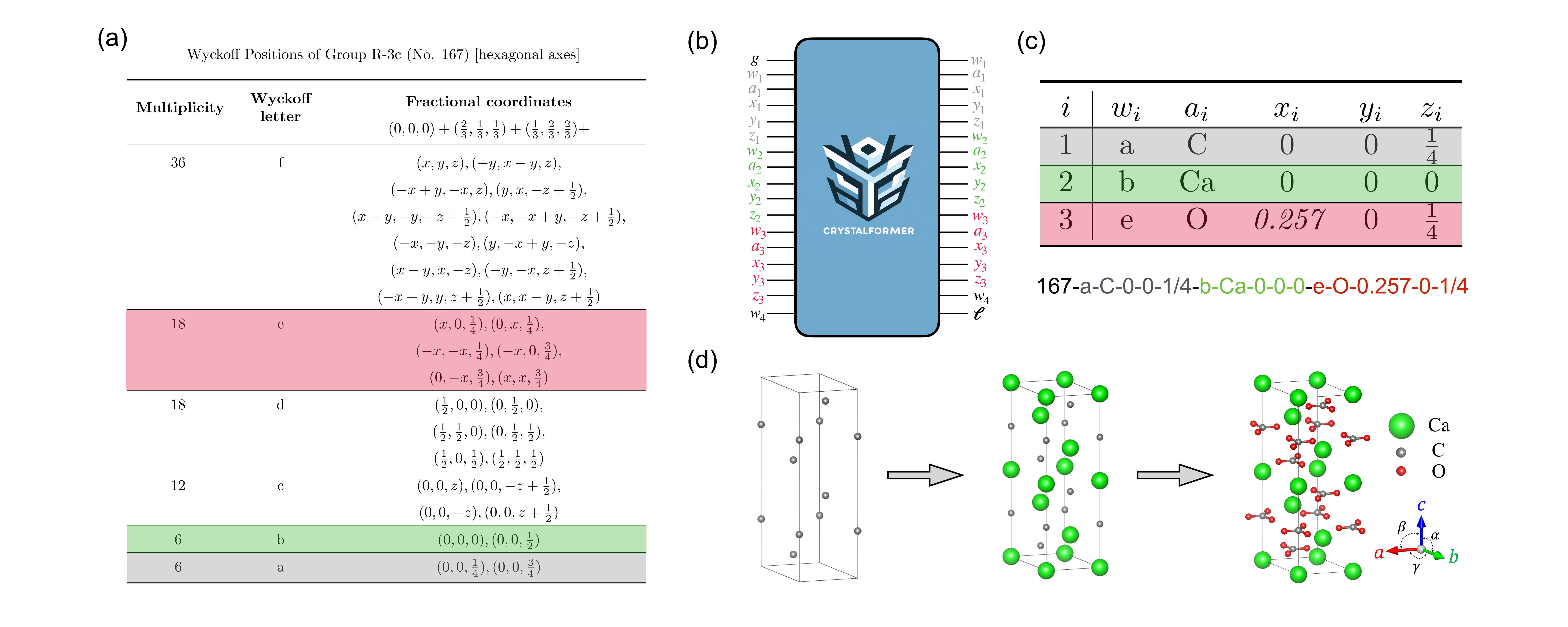}
    \caption{
    (a) The Wyckoff positions of the $R\bar{3}c$ space group (No. 167).  
     We highlight the occupied Wyckoff positions of calcite CaCO$_3$ crystal which belongs to this space group. Carbon, calcium, and oxygen atoms occupy the `6a', `6b', and  `18e' positions, respectively.   
    (b) The \texttt{CrystalFormer} is a decoder-only autoregressive transformer that models the space group controlled crystal structures by predicting probabilities of the Wyckoff letter $w_i$, chemical element $a_i$, and fractional coordinates $(x_i, y_i, z_i)$ of each symmetry inequivalent atom, and finally, the lattice parametrized by $\boldsymbol{\ell}$ sequentially. 
    (c) The crystal data of CaCO$_3$ is summarized in a table. 
    In the table, the x-coordinate of oxygen atom $x_3=0.257$ is the only continuous variable that needs to be predicted. All other fractional coordinates are fixed by discrete data like the space group number and Wyckoff letters. The string below the table shows the sequential representation of the CaCO$_3$ crystal with space group, Wyckoff letter, and atom species as the input to the \texttt{CrystalFormer} model. 
    (d) Autoregressive generation of the crystal. One first places carbon atoms at the `6a' position, then places calcium atoms at `6b' position, and finally places oxygen atoms at `18e' position. In each step of the sampling procedure, there is a choice of the Wyckoff positions, atom species, and the fractional coordinates if they are still unspecified.
    }
    \label{fig:calcite_and_model}
\end{figure*}

Space group symmetry imposes significant constraints on a crystal. First of all, the space group identifies the crystal system to which a crystal belongs, thereby limiting the permissible values for the lattice parameters that define the length and angles of the crystal's unit cell. Moreover, the symmetry operations associated with a given space group ensure that symmetry equivalent atoms are consistently mapped among themselves in the crystal. This requirement enforces strict conditions regarding the types of chemical elements present, their specific locations within the crystal, and the number of each chemical species in the unit cell. A key concept to express these constraints is the Wyckoff positions, which delineate unique areas within a unit cell that are defined by the symmetry operations of the crystal's space group. These positions are represented as fractional coordinates, enabling precise definition relative to the unit cell's axes. For example, Fig.~\ref{fig:calcite_and_model}a shows the Wyckoff positions for the space group $R\bar{3}c$ (No. 167). The Wyckoff positions are labeled by letters in the alphabet, starting from special points in the bottom to general positions in the top. The multiplicity counts the number of equivalent positions connected by the space group symmetry operations. All of them should be occupied by the same type of atoms to uphold the space group symmetry.  For example, the top row of the table in Fig.~\ref{fig:calcite_and_model}a contains the general position $(x,y,z)$ that can be mapped to 36 positions under the symmetry operations of the $R\bar{3}c$ space group. 

Nature tends to place atoms in those special Wyckoff positions at the bottom of the table. For example, we highlight the occupied Wyckoff positions of calcite (CaCO$_3$) crystal in Fig.~\ref{fig:calcite_and_model}, associated with the $R\bar{3}c$ space group. One sees that the Wyckoff letter `6a' and `6b' deterministically define the locations of the carbon and calcium atoms within the unit cell. In addition, it follows that  $a = b$, and $\alpha = \beta = 90 ^{\circ}, \gamma = 120^{\circ}$ as the $R\bar{3}c$ space group belongs to the trigonal crystal system. Ultimately, despite having 30 atoms in the unit cell, there are only three continuous degrees of freedom for the CaCO$_3$ structure: the x-coordinate of oxygen atom $0.257$ and the lattice constants $a=b=4.99 \AA$ and $c = 17.07 \AA$. All other information about the crystal structure can be specified via discrete data such as the Wyckoff letters and chemical species. 

The prominent discrete and sequential features illustrated in Fig.~\ref{fig:calcite_and_model} are ubiquitous in crystalline materials. The Wyckoff positions not only specify possible locations of atoms in the unit cell, but their associated multiplicities also put strong constraints on the number of atoms. Therefore, space group symmetry significantly reduces the degrees of freedom of crystalline materials. 
Failing to exploit this information in generative modeling not only renders learning inefficient, it also severely impairs the generalization ability of the model. For example, the performance of the generative model quickly deteriorates as the number of atoms increases due to it is challenging to generate highly symmetric crystal structures~\cite{zeni2023mattergen}. On the other hand, statistical analysis shows that the Wyckoff sequences of known inorganic compounds ~\cite{hornfeck2022combinatorics} are far from being exhausted, implying there are statistical correlations to be exploited to compress the materials database. 

In this paper, we introduce \texttt{CrystalFormer}, an autoregressive transformer for generative modeling of crystalline materials. \texttt{CrystalFormer} models the joint probability distribution of Wyckoff positions, chemical species, and lattice parameters of crystals with a given space group. By treating the Wyckoff positions as the first class citizen in the model, \texttt{CrystalFormer} seamlessly and rigorously integrates the space group symmetry into crystal probabilistic modeling. As shown in Table~\ref{tab:count}, explicit modeling of the Wyckoff positions greatly reduces the space of crystalline materials. The space group-informed transformer exploits this fundamental inductive bias to greatly simplify the learning and generation of crystals. 


\section{Method}

We will first introduce the \texttt{CrystalFormer} model, then reveal the chemical intuition encoded in the trained model by inspecting generated crystal samples. These inspections also build up understandings of the strength of the model. 

\subsection{\texttt{CrystalFormer}}
\label{sec:crystalformer}
We will introduce the design, training, and sampling of the \texttt{CrystalFormer} model.

\subsubsection{Model}

To exploit the space group symmetry of the crystal, we focus on the Wyckoff positions of symmetry-inequivalent atoms. 
Wyckoff letters follow the alphabetical order, where ``a''
stands for the positions with the highest order of site symmetry for the given space group. Later letters in the alphabet indicate more general positions with reduced site symmetries.  
Note that the information of the space group number and Wyckoff letter fully determine the multiplicities of the Wyckoff positions. In cases where the atom positions are not fully fixed by the Wyckoff letter, we will also consider the remaining fractional coordinates, e.g. the $x$-coordinate of the oxygen atoms in the CaCO$_3$ example shown in Fig.~\ref{fig:calcite_and_model}.
To generate crystals, one samples the Wyckoff letter, chemical element, and fractional coordinates of each atom sequentially. The sampling procedure starts from special higher symmetry sites with smaller multiplicities and then goes on to general lower symmetry sites with larger multiplicities. 

With these considerations, we define a crystal data as $\boldsymbol{\mathcal{C}} = \{\boldsymbol{W}, \boldsymbol{A}, \boldsymbol{X}, \boldsymbol{L}\}$. Here $\boldsymbol{W} = [ w_1, w_2, \ldots, w_n ]$ are  Wyckoff letters and $\boldsymbol{A} = [ a_1, a_2, \ldots, a_n ] $ are chemical species. Here, $n$ stands for the number of symmetrically inequivalent atoms in the conventional unit cell. For example, as shown in Fig.~\ref{fig:calcite_and_model}b one has $n=3$ for CaCO$_3$.  Explicitly including the Wyckoff letter in the generative modeling is the key of the present work. Next, 
$\boldsymbol{X} =[ (x_i, y_i, z_i) ] \in \mathbb{R}^{n\times 3}$
are the fractional coordinates of symmetrically inequivalent atoms. 
Lastly, $\boldsymbol{L}=[a,b,c,\alpha,\beta,\gamma]$ denotes the lattice parameters of the conventional unit cell of the crystal.

The central quantity to focus on is the conditional probability of a crystal $\boldsymbol{\mathcal{C}}$ given the space group number $g\in [1, 230]$:  
$p(\boldsymbol{\mathcal{C}}|g)$. Since the space group is a fundamental characterization for crystalline materials, $g$ is a key control variable that greatly simplifies the distribution over the entire crystal materials space. In practical applications of crystal structure prediction and material design, the space group can either be considered separately as a control variable or predicted based on material composition \cite{zhao2020machine,PhysRevMaterials.4.123802,wang2023crystallographic,venkatraman2024accurate}. 

We express the space group conditioned probability distribution of crystals as an autoregressive product of conditional probabilities
\begin{eqnarray}
p(\boldsymbol{\mathcal{C}}|g)  &=  & p(w_1 | g) \times   \nonumber \\  
 & & p(a_1 | g, w_1) \times \nonumber \\
 & & p( {x}_1 | g, w_1, a_1 )\times \nonumber \\  
 & & p( y_1 | g, w_1, a_1, x_1 )\times \nonumber \\   
 & & p( z_1 | g, w_1, a_1, x_1, y_1 ) \times  \cdots \times  \nonumber  \\   
 & &  p( \boldsymbol{L} | g, w_1, a_1, x_1, y_1, z_1 \ldots, w_n, a_n, x_n, y_n, z_n).
 \label{eq:pcg}
\end{eqnarray} 
At first sight, it may appear unnatural to employ an autoregressive model for crystals since there seems to be no obvious order for atoms in the unit cell. However, the sequential nature of Wyckoff positions suggests a natural way to arrange symmetrically inequivalent atoms in an alphabetical order of the Wyckoff letters. Following this key observation, we represent crystal data as sequences of space groups, Wyckoff letters, chemical species, and fractional coordinates of each symmetrically inequivalent atom. Together with the information lattice parameters, such sequence fully characterizes the compositional and structural information of crystalline material. Since statistical analysis reveals that anions are in less symmetric positions than cations for inorganic crystals~\cite{urusov2009frequency}, one would expect that anion atoms will typically appear after cation atoms in such a sequence. For example, CaCO$_3$ is represented as a string ``167-a-C-0-0-1/4-b-Ca-0-0-0-e-O-0.257-0-1/4''. Autoregressive sampling of such a string means the model generates the crystal by placing the atoms sequentially into the unit cell, starting from the special position with high site symmetry to the general position with the lowest site symmetry, see Fig.~\ref{fig:calcite_and_model}d.

We model the conditional probability of the Wyckoff letters $\boldsymbol{W}$ and chemical species $\boldsymbol{A}$ as categorical distributions. 
On the other hand, we model the conditional probability of the factional coordinates $\boldsymbol{X}$ as a mixture of von Mises distribution for continuous periodic variables. For Wyckoff positions with multiplicities greater than one, we only consider the first of fractional coordinates that appear in the international tables for crystallography~\cite{hahn1983international}. Lastly, we model the conditioned distribution of lattice parameters as a Gaussian mixture model. 

We build \texttt{CrystalFormer}, an autoregressive transformer~\cite{vaswani2017attention} to model the space group conditioned-probability distribution of crystalline materials~\Eq{eq:pcg}. 
The space group number $g$ is the first input to \texttt{CrystalFormer}. The remaining inputs are the Wyckoff letter, chemical species, and fractional coordinates of each atom. One can go through the table of Fig.~\ref{fig:calcite_and_model}b in a raster order to collect these atomistic features. We feed vector embeddings of the space group number, Wyckoff letter, and the chemical species input to the \texttt{CrystalFormer}. In particular, we also concatenate the vector embedding of $g$ to all other inputs since it is the key control variable for the crystal generation. Moreover, we have also provided the multiplicity of each Wyckoff position as an additional feature. The multiplicity can be easily inferred from the space group and the Wyckoff letters. We feed the fractional coordinates as Fourier features into the transformer so that the model preserves the periodicity of the unit cell~\cite{wirnsberger2020targeted,jiao2023crystal}. 
We pad the atom sequence up to a maximum length and treat the output as parameters of the conditional probability distribution \Eq{eq:pcg}, see Fig.~\ref{fig:calcite_and_model}b. At the location of the first padding atom, we predict the lattice parameters. 

We implement a number of constraints in the model to further reduce its phase space. First, the Wyckoff letters should be valid for the given space group. For example, for the space group $R\bar{3}c$ (No. 167) the Wyckoff letters go from `a' up to `f'. Second, we require that the Wyckoff letters $w_i$ follow alphabetical order in the sequence~\footnote{We follow the convention that capital letters appear \emph{after} lower case letters. This handles the edge case of the $Pmmm$  space group (No. 47) whose Wyckoff letters used up 26 lowercase letters and reached `A' for the generic position. In addition, we use the letter `X' to indicate the Wyckoff position of padding atoms that appear at the end of the sequence, see e.g. $w_4$ of Fig.~\ref{fig:calcite_and_model}b.}. Lastly, the Wyckoff positions with no free fractional coordinates (such as `a', `b', and `d' positions in the $R\bar{3}c$ space group) can only be occupied once. Those constraints are implemented by setting the logit biases of Wyckoff letters to mask out invalid sequences ~\cite{openai_logit_bias,xie2023m}. 

The design of \texttt{CrystalFormer} focuses mostly on the space group symmetries which we believe to be the most important inductive bias for crystalline materials. This design decision significantly impacts the treatment of other symmetries. First, it is often possible to place the origin of the unit cell at the inversion center of the specified space group. The chosen origin naturally fixes the continuous translation invariance of fractional coordinates. 
Second, by only considering symmetry-inequivalent atoms and labeling them with Wyckoff letters, one fixes most of the permutation invariance over atom of the same type in the representation. 
For those Wyckoff positions with continuous degrees of freedom, there may be multiple symmetry-inequivalent atoms with the same Wyckoff letters. We arrange these atoms according to the lexicographic order of fractional coordinates~\cite{parthe1984standardization} in the sequence. Note that in a crystal environment, the same type of atoms occupying different Wyckoff positions could be regarded as distinguished particles as they generally have different site symmetry. 
Lastly, the periodicity of the fractional coordinates is respected in \texttt{CrystalFormer} since they are treated as periodic variables following the von Mises distribution. 

\subsubsection{Training}

The \texttt{CrystalFormer} is trained by minimizing the negative log-likelihood 
\begin{equation}
\mathcal {L} =  -\mathop{\mathbb{E}}_{\boldsymbol{\mathcal{C}}, g} \left [  \ln p(\boldsymbol{\mathcal{C}} |g) \right ],  
\end{equation}
where the structures $\mathcal{C}$ and the corrosponding space group $g$ of crystals are sampled from the training dataset. Writing out $p(\boldsymbol{\mathcal{C}} |g) $ according to \Eq{eq:pcg}, the objective function contains the negative log-likelihood of discrete variables such as Wyckoff letters $\boldsymbol{W}$ and chemical species $\boldsymbol{A}$, as well as continuous variables such as fractional coordinates $\boldsymbol{X}$ and the lattice parameters $\boldsymbol{L}$.  For the continuous variables $\boldsymbol{X}, \boldsymbol{L}$ in the
 objective function, we consider only active ones that are not fixed by the space groups and Wyckoff letters. In this way, those special fractional coordinates (e.g. $0$, $\frac{1}{4}$) and lattice parameters (e.g. $90^\circ$, $120^\circ$) which were already fixed by the chosen space group and Wyckoff letter will not not contribute to the loss function. 

\begin{figure}[t!]
    \centering
    \includegraphics[width=1\linewidth]{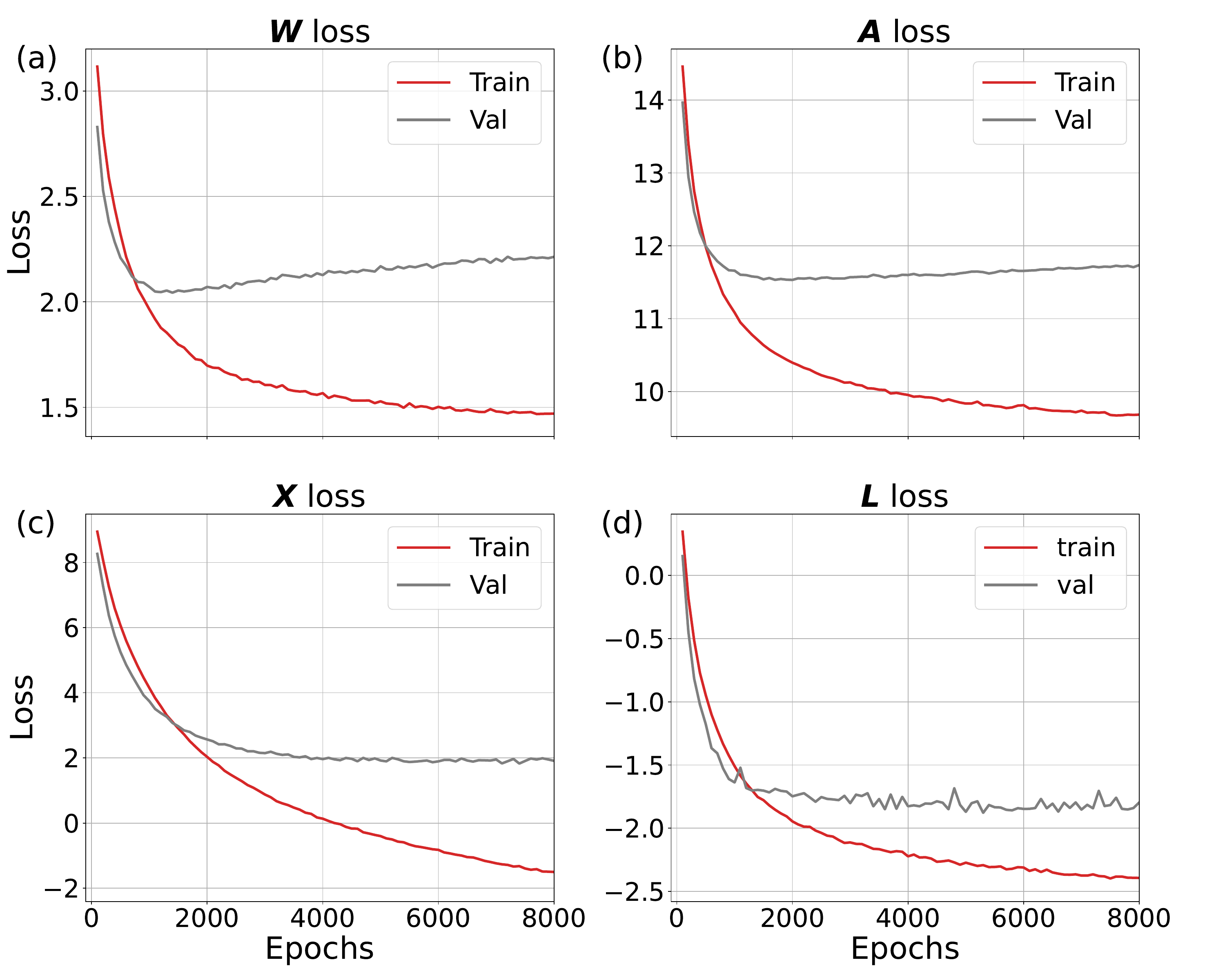}
    \caption{Break up of the training and validation losses for (a) Wyckoff letters, (b)  chemical species, (c) fractional coordinates, and (d) lattice parameters over training epochs. 
    }
    \label{fig:lcurve}
\end{figure}

In the present work, we train the \texttt{CrystalFormer} using the MP-20 dataset~\cite{xie2021crystal}. MP-20 is a popular dataset that represents a majority of experimentally known crystalline materials at ambient conditions with no more than 20 atoms in the primitive unit cell. The training dataset contains 27136 crystal structures.  The subdivision of the training samples according to the space group has greatly reduced the number of samples in each space group category. On top of that, the distribution of training samples is quite uneven among the space groups, which reflects the imbalanced distribution of crystals over space groups in nature~\cite{urusov2009frequency}. In fact, there is no training data in 61 out of 230 space groups as shown in Fig.~\ref{fig:valid_hist}. Nevertheless, we still employ the MP-20 as the training set so that the performance of the model can be more easily gauged with the others in the literature, see appendix~\ref{sec:validity}. 
Note that the \texttt{CrystalFormer} can generate reasonable samples even for those space groups without any training data. This because the model can exploit knowledge learned from other space groups to place suitable atoms in the Wyckoff positions due to weight sharing. Moreover, since the sampling process makes use the of Wyckoff position table. The three dimensional coordinates of atoms are not completely random even for unseen space groups.
Fig.~\ref{fig:lcurve} shows a breakup of the learning curves for the Wyckoff position, chemical species, fractional coordinates, and lattice parameters. We select the model checkpoint with the lowest total validation loss to generate crystal samples.

\subsubsection{Sampling}
\label{sec:sampling}

To sample crystals from the \texttt{CrystalFormer}, one needs to specify a space group number and a list of possible chemical elements. The \texttt{CrystalFormer} samples the atoms one by one, starting from more symmetric specific positions with lower multiplicities till less symmetric general positions with larger multiplicities. We use the information of the space group and Wyckoff letter to control the sampling of fractional coordinates. By applying the symmetry projection to the sampled fractional coordinate, one rectifies it and ensures the generated fractional coordinates are compatible with the Wyckoff positions. One can also mask out the logits of chemical species so that only a number of selected elements will be sampled. The number of symmetrically inequivalent atoms may fluctuate in the sampling procedure. Once one has sampled a padding atom, the model predicts the lattice parameters under the space group constraint. Moreover, we introduce a temperature parameter $T$ in the sample distribution $p(\boldsymbol{\mathcal{C}} |g )^{1/T}$. With $T<1$ we will draw samples from a sharper distribution, while $T>1$ gives more diversity in the generated samples. In the present paper, we will generate crystals using temperature $T=1$ unless mentioned explicitly. 

Besides autoregressive sampling, one can also perform Markov chain Monte Carlo (MCMC) sampling based on the likelihood \Eq{eq:pcg} of the \texttt{CrystalFormer}. MCMC sampling can walk through the crystalline materials space starting from an existing crystal structure. At each step of the random walk, one proposes a configuratoin update in terms of element substitution, atom position shift, or lattice deformation to change the crystal from $\boldsymbol{\mathcal{C}}$ to $\boldsymbol{\mathcal{C}}^\prime$, then accepts or rejects the proposal according to the model probability following the Metropolis acceptance rule $\min \left[ 1, \frac{p(\boldsymbol{\mathcal{C}}^\prime|g)}{ p (\boldsymbol{\mathcal{C}}|g)} \right]$. MCMC sampling is particularly useful for incorporating additional constraints or guidance in the sampling procedure. Moreover, during the burn-in phase of such MCMC sampling, the generated samples will be similar to the starting material, which may be a desired feature in certain cases.

\subsection{\texttt{CrystalFormer} learns chemical intuition by compressing materials database}
\label{sec:intuition}

Nature favors symmetrical crystal structures. Crystallographic space groups quantify this inductive bias of nature, thereby significantly simplifying the spaces of crystal materials. In light of the space group symmetries, crystals also have an unexpected yet natural sequential and discrete representation, which derives from two tables in nature: the periodic table of elements determined by quantum mechanics and the table of Wyckoff positions of the 230 space groups determined by group theory. To construct a certain crystal, we only need to select atoms from the periodic table and place them sequentially into the Wyckoff positions in the unit cell. In this crystal language, the “word order” is determined by the alphabetical order of Wyckoff letters, the “grammar” corresponds to the solid-state chemistry rules, and the “synonyms” represent interchangeable elements (Sec.~\ref{sec:aa correlation}), the “sentence length” correspond to atom number in the unit cell (Sec.~\ref{sec:length}), 
and the “idioms” correspond to common chemical coordination (Sec.~\ref{sec:wa correlation}).

\texttt{CrystalFormer} employs an autoregressive transformer to learn the crystal language, thereby exploring yet-to-be-discovered crystalline materials. It compresses and internalizes the crystal materials database, expressing 
solid-state chemical knowledge through neural network parameters; reflecting the associative ability of material space through neural network activations; and describing chemical intuition through the model probability (Sec.~\ref{sec:likelihood}). Similar to generative models used for generating text, images, and videos, \texttt{CrystalFormer} can directly generate “realistic” crystal materials. However, rather than worrying about the fake contents of AI-generated media, these AI-generated crystal materials could potentially be synthesized and be useful to human civilization.

Next, we will inspect the learned features and sample statistics of the model to build up an understanding of the \texttt{CrystalFormer}. We carry out inspectations for a few selected space groups. The findings are neverthelss general. These findings  provide understandings and confidence of the model, therefore direct us to the suitable applications of \texttt{CrystalFormer}.  

\subsubsection{Atom embeddings and chemical similarity}
\label{sec:aa correlation}

\begin{figure}[t!]
    \centering
    \includegraphics[width=1\linewidth]{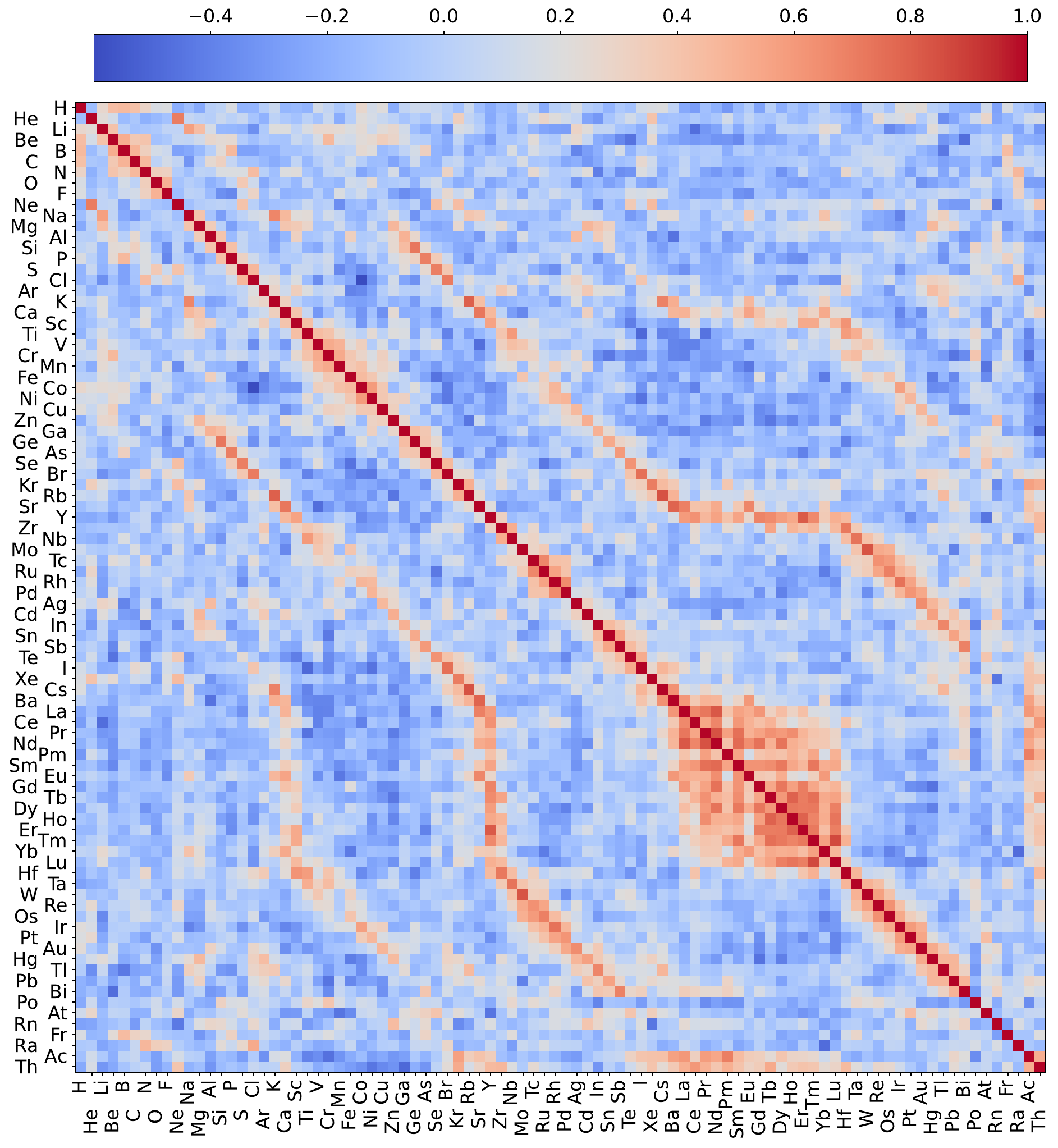}
    \caption{The cosine similarity matrix for the chemical species based on the learned vector embeddings. The reddish color suggests similar chemical elements in the crystal environment. 
    }
    \label{fig:aa}
\end{figure}

Fig.~\ref{fig:aa} visualizes the cosine similarity of the learned vector embedding of the chemical species. Red colors in the figure indicate similar chemical species identified by the model. One sees the chemical similarity within groups of elements show up as off-diagonal red stripes. Moreover, there are visible clusters for Lanthanide elements (La-Lu). The plot also suggests the similarity between the lanthanides and other rare-earth elements (Y and Sc). The features shown in Fig.~\ref{fig:aa} are strikingly similar to the similarity map constructed purposely based on substitution pattern~\cite{hautier2011data,glawe2016optimal} which was later used for substitution-based material discoveries~\cite{wang2021predicting,merchant2023scaling}. In the context of language modeling, the chemical similarities correspond to synonyms of chemical species tokens. Having the ability to learn chemical similarities from data
~\cite{hautier2011data,glawe2016optimal,jain2018atomic,zhou2018learning, chen2019graph, zhang2022dpa,wang2022crabnet,antunes2023crystal} is an encouraging signal that the model is picking up atomic physics for generating reasonable crystal structures with maximum likelihood based training. 

\subsubsection{Atom number distributions}
\label{sec:length}
\begin{figure}[t!]
    \centering
    \includegraphics[width=1\linewidth]{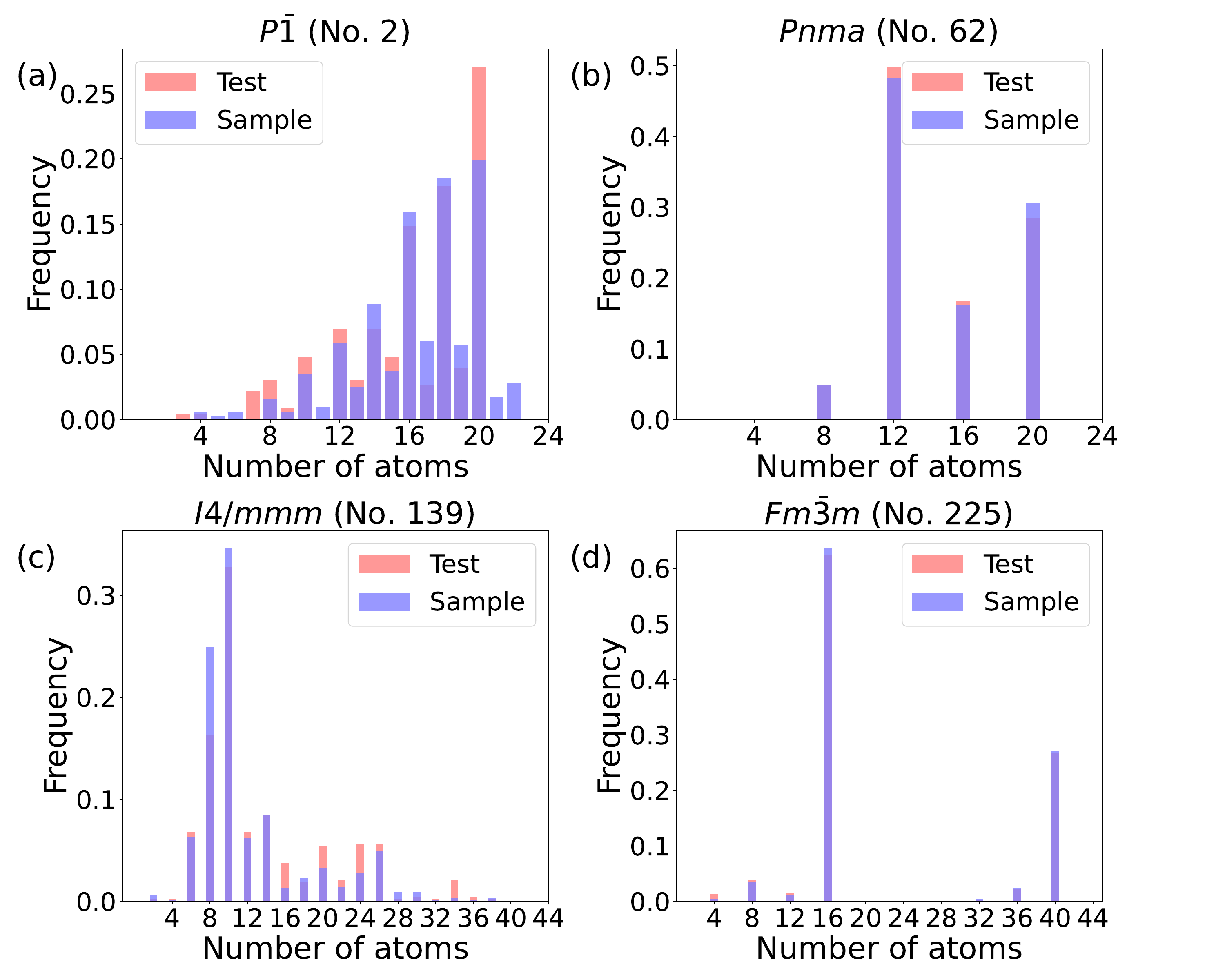}
    \caption{The histogram for \emph{total number} of atoms in the unit cell for (a) $P\bar{1}$ (b) $Pnma$ (c) $I4/mmm$ (d) $Fm\bar{3}m$ space groups in the test dataset and in the generated samples. 
     }
    \label{fig:atom_hist}
\end{figure}

The number of atoms in the unit cell corresponds to the length of non-padding atoms in the crystal string representation, which is captured well by \texttt{CrystalFormer}. 
Fig.~\ref{fig:atom_hist} presents the histogram of the total number of atoms in the conventional unit cell for several space groups. One sees a nice agreement between the atom number distribution in the test dataset and the generated samples. 
In addition, it appears that space group $g$ is the key latent variable that decomposes the multi-modal atom number distribution of crystals. This is understandable because the number of atoms is determined by the sum of the multiplicities of occupied Wyckoff positions. Therefore, the space group symmetry is a key control variable for the atom number distribution. Incorporating Wyckoff positions information into the \texttt{CrystalFormer} model architecture removes the necessity of querying the training data to find out the number of atoms for a targeted space group~\cite{zeni2023mattergen} during generation. 

\begin{figure*}[t!]
    \centering
    \includegraphics[width=1\linewidth]{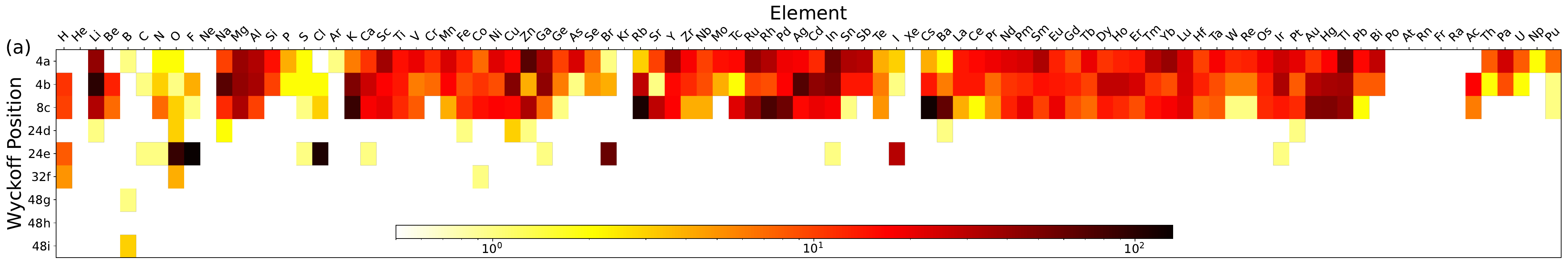}
    \includegraphics[width=1\linewidth]{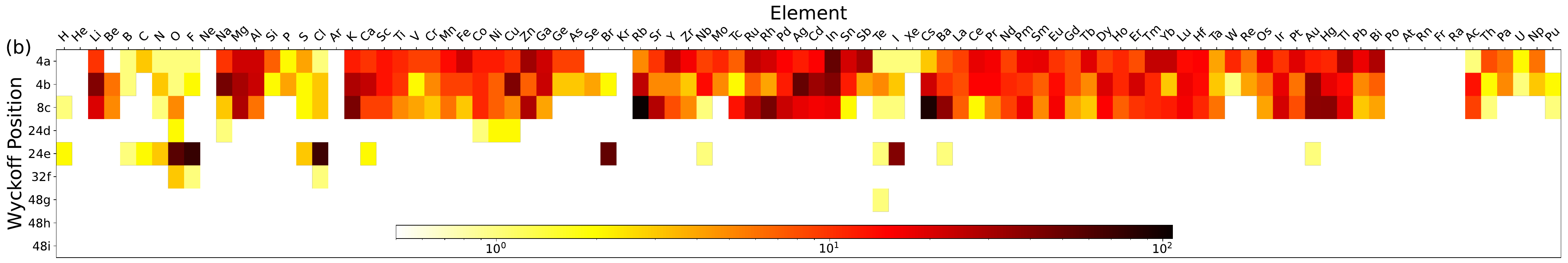}
    \caption{The heat map for Wyckoff positions and atom species of (a) the test dataset and (b) generated samples for the $Fm\bar{3}m$ space group (No. 225). It is an analog of bigram frequency statistics of language modeling, which shows the prefereed Wyckoff positions of different chemical species. 
    }
    \label{fig:aw}
\end{figure*}

Recently, Ref.~\cite{gazzarrini2024rule} reports an abundance of inorganic compounds whose primitive unit cell contains a number of atoms that is a multiple of four. There are different ways to reason about the observed ``rule of four'' depending on one's view of how a crystal is formed. For example, one can often break inorganic solids into polyhedra as building blocks. Otherwise, Ref.~\cite{palgrave2024explanation} considers the most probable values of 
the number of atoms in a formula unit and the number of formula units per primitive cell. In line with the discussion here, the ``rule of four'' is the combination of three factors 1) the distribution of crystalline materials among space groups~\cite{urusov2009frequency}; 2) the distribution of atoms in Wyckoff positions~\cite{hornfeck2022combinatorics} of a given space group; and 3) the multiplicities of Wyckoff positions and multiplicities of conventional versus primitive cells. The first two are statistical rules determined by the inter-atomic interactions while the third one is a mathematical fact of space group theory. In the end, the point we want to make is that the design of \texttt{CrystalFormer} and its associated crystal representation allow it to learn the ``rule of four'' and many other to-be-discovered ``rules'', which manifest themselves as marginal statistics of learned probability distribution. Most importantly, \texttt{CrystalFormer} will utilize these ``empirical rules'' when generating novel yet reasonable crystal samples.  

\subsubsection{Wyckoff-Atom gram}
\label{sec:wa correlation}

Fig.~\ref{fig:aw} shows heat maps of Wyckoff positions and chemical species for the  $Fm\bar{3}m$  space group (No. 225). The heat map is analogous to bigram frequency statistics in language modeling. In the present context, it reveals interesting solid-state chemistry knowledge related to where each atom tends to appear in a unit cell.
First of all, one sees that most atoms occupy special Wyckoff positions (Wyckoff letters at the beginning of the alphabet) with higher site symmetries. The distribution of generated data is in agreement with test data and recent statistics~\cite{hornfeck2022combinatorics}. Moreover, there are vertical blanks at the locations of inert elements (He, Ne, Ar...) as they are rare in crystalline materials. Lastly, one sees that oxygen and halogen elements (F, Cl, Br, I) appear quite often in the Wyckoff position ``24e'', which means these high electronegative elements form polyhedra enviorment for other atoms~\cite{urusov2009frequency}. Overall, we see the \texttt{CrystalFormer} has learned these key motifs for generating crystalline materials. On the other hand, one also observes that several Wyckoff locations of the hydrogen are missing in the generated samples compared to the test dataset. We believe that is due to that the hydrogen element takes only about $0.4\%$ in the training data for the $Fm\bar{3}m$ space group. Collecting more data with better coverage of elements will be crucial to further boost the performance of the current model.  

Along the same line of thoughts, coordination polyhedra~\cite{alvarez2005polyhedra} and lattice structure~\cite{regnault2022catalogue} manifest themselves as higher-order n-gram correlations of Wyckoff position and atom species in the crystal language,  which will be captured by the \texttt{CrystalFormer}. There has been a long history of mining empirical chemistry rules encoded in materials data and then using them to instruct the search of crystal structures~\cite{pauling1929,goldschmidt1929crystal,pettifor1988structure, fischer2006predicting, glawe2016optimal, allahyari2020coevolutionary}. Our analysis shows that \texttt{CrystalFormer} ingests chemical intuition, be it speakable or unspeakable, in the training data for generating new materials. 

\subsubsection{Crystal likelihoods}
\label{sec:likelihood}

\begin{figure*}[ht!]
    \centering
    \includegraphics[width=0.8\linewidth]{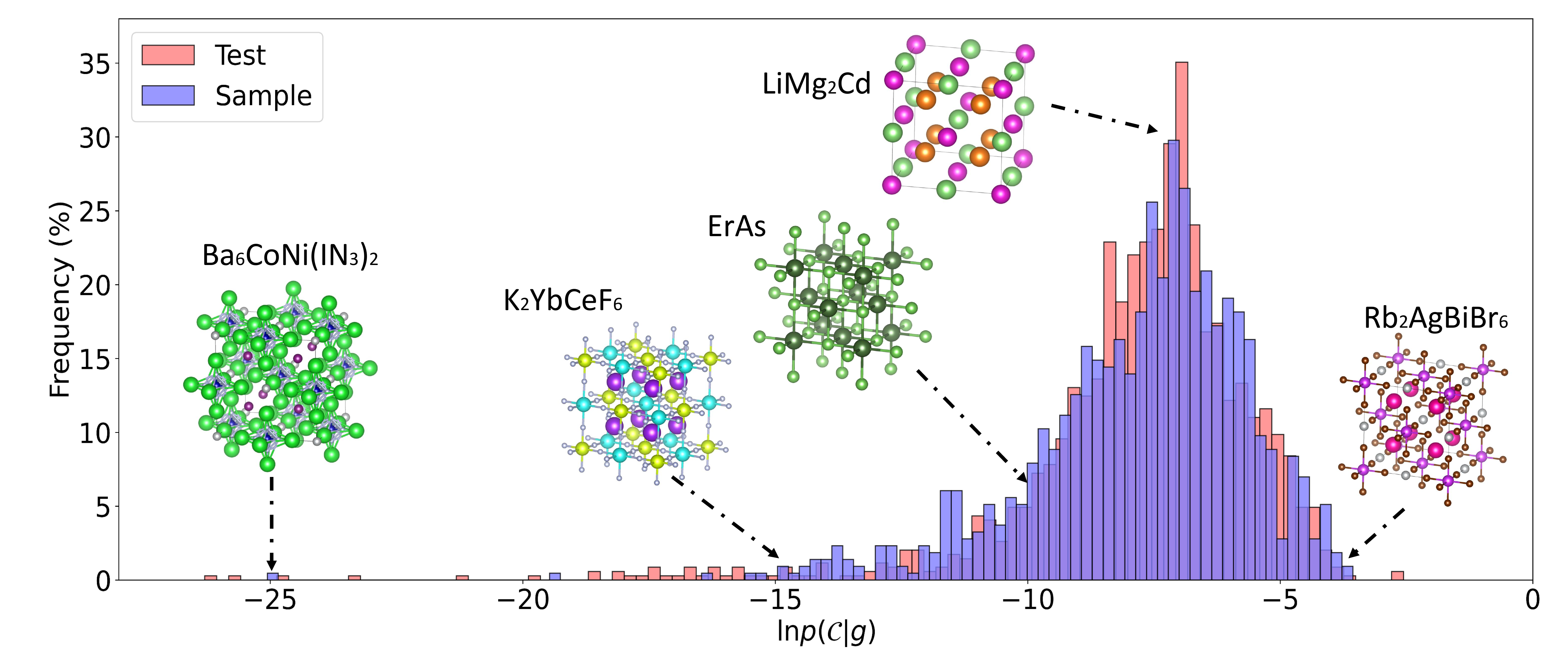}
    \caption{ 
     The histogram of log-likelihoods of 1000 samples in the $Fm\bar{3}m$ (No. 225) space group and the test dataset. 
    The insets visualize the crystal structure of a few generated samples. Rb$_2$AgBiBr$_6$ and LiMg$_2$Cd are in the training dataset. ErAs is in the validation dataset. K$_2$YbCeF$_6$ and Ba$_6$CoNi(IN$_3$)$_2$ are not in the MP-20 or Materials Project database. 
    \label{fig:logp_hist}
   }
\end{figure*}

\texttt{CrystalFormer} compresses chemistry knowledge stored in the material dataset into its parameters. In addition to generating crystal samples, \texttt{CrystalFormer} can also compute the likelihoods of crystals via \Eq{eq:pcg}. Therefore, it is also possible to employ \texttt{CrystalFormer} in likelihood-based Monte Carlo search besides sampling crystals directly.

Fig.~\ref{fig:logp_hist} shows the agreement of the likelihoods of generated samples and samples in the test dataset. We also visualize structures of a few generated samples which are deemed to be very likely, typical, and unlikely according to their likelihood values. We have checked that likelihood is related to the energy of the crystal by locally perturbing the fractional coordinates and lattice parameters. However, we did not observe a correlation between the likelihood of these crystals and their energies on a global scale. We envision the landscape of likelihoods is much less rough compared to the potential energy surface of crystalline materials. Intuitively, it means that the \texttt{Crystalformer} compresses the materials space into a more compact space without many holes that correspond to infeasible high energy states. Therefore, likelihood-based exploration of the crystal space discussed in Sec.~\ref{sec:sampling} can be more efficient compared to traditional sampling approaches based on the Boltzmann distribution based on physical energy functions. 

\section{Results}
\label{sec:application}

We now move on to the practical applications of \texttt{CrystalFormer} to materials discovery and design. 
Compared to many existing materials generation models, \texttt{CrystalFormer} offers precise control over space group symmetry and enables efficient computation of model likelihood. These unique features open a wide range of possibilities for integrating it with existing computational software and machine-learning models in a flexible way as we demonstrate below. For these applications, we have excluded radioactive elements from the samples~\cite{cheetham2024artificial}. 

\subsection{Symmetry-conditioned random structure initialization}
\label{sec:initialization}


Crystal structure prediction has long been the dream of solid-state chemistry and computational material science researchers~\cite{maddox1988crystals}. Typical crystal structure prediction workflow consists of two steps. First, one randomly initializes a batch of diverse crystal structures as candidates. Second, one optimizes the crystal structures via local and global optimization strategies. Utilizing space group symmetries plays a crucial role in both steps, as symmetry enlarges the span of the energy distribution~\cite{wales1998symmetry, avery2017randspg, pyxtal} and reduces the search space. 

It is a common practice for crystal structure prediction software~\cite{PhysRevB.82.094116,pickard2011ab,lyakhov2013new,avery2017randspg, falls2020xtalopt, pyxtal} and structure search~\cite{cheng2022crystal, wang2023symmetry, zhang2024symmetry} to exploit space group symmetry in the crystal structure initialization. 
However, such an initialization approach faces combinatorial difficulty as the number of chemical species and atoms in the unit cell grows. The \texttt{CrystalFormer} is ready to act as a drop-in replacement of random structure initialization for crystal structure prediction. In this way, one bypasses the curse-of-dimensionality of exact enumeration~\cite{avery2017randspg} with a data-driven probabilistic approach. Moreover, the ability of \texttt{CrystalFormer} to generate diverse and near-stable structures can greatly reduce the computational costs of downstream optimizations. 

We select seven space groups $P\bar{1}$ (No. 2), $C2/m$ (No. 12), $Pnma$ (No. 62), $I4/mmm$ (No. 139), $R\bar{3}m$ (No. 166), $P6_{3}/mmc$ (No. 194), and $Fm\bar{3}m$ (No. 225) as representatives of the seven crystal systems. We randomly generate 100 crystals for each space group using \texttt{CrystalFormer}. On the other hand, we employ PyXtal ~\cite{pyxtal} to generate crystal samples with the same stoichiometry in the same space groups. We then carry out structure relaxation using density functional (DFT) calculations. 

\begin{figure*}[t!]
    \centering
    \includegraphics[width=1\linewidth]{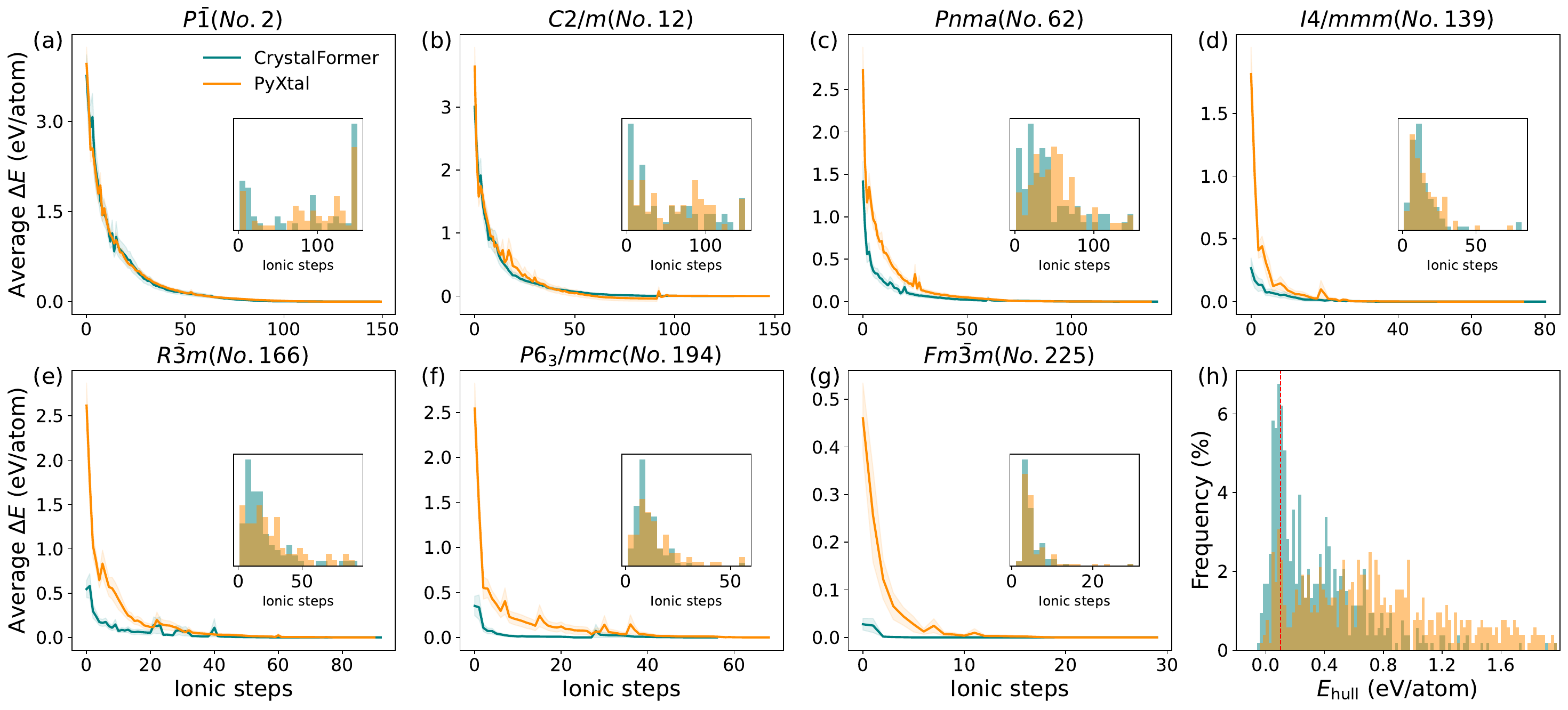}
    \caption{(a)--(g) Average energy difference versus relaxation steps for seven representative space groups. The insets show the distribution of ionic steps. (h) The histograms of energy above the convex hull for relaxed crystal structures. The dashed line indicates the criterion for selecting candidates for stable materials listed in Appendex~\ref{sec:initialization-app} since materials with $E_\mathrm{hull} < 0.1$ eV/atom are usually metastable and have the potential to be synthesized~\cite{sun2016thermodynamic}.  
    }
    \label{fig:initialization}
\end{figure*} 

\begin{table*}[t]
\resizebox{2\columnwidth}{!}{
\begin{threeparttable}
\centering
\caption{For each space group we randomly generate 100 crystal structures with the same composition using \texttt{CrystalFormer} and PyXtal. We carry out energy relaxation using DFT calculations and report the number of converged samples,  the number of structures that maintain the original space group symmetry, the average RMSD between generated and relaxed structures, and the averaged energy above the convex hull.
}
\renewcommand{\arraystretch}{1.2}
\begin{tabular}{llcccccccccc}
\toprule
\multirow{2}{*}{Space group}  & \multirow{2}{*}{Crystal system} & \multicolumn{2}{c}{{Converged structures $\uparrow$}} &   \multicolumn{2}{c}{{Retain symmetry   $\uparrow$}} &   \multicolumn{2}{c}{{RMSD \tnote{1} ($\AA$)   $\downarrow$}} 
& \multicolumn{2}{c}{{$E_\mathrm{hull}$ \tnote{1} (eV/atom)   $\downarrow$}}\\
    &                   &  \texttt{CrystalFormer}        & PyXtal    &   \texttt{CrystalFormer}        & PyXtal   &   \texttt{CrystalFormer}          & PyXtal   &   \texttt{CrystalFormer}          & PyXtal       \\ 
\midrule
$P\bar{1}$ (No. 2) & Triclinic       &    46     & \textbf{67}     & 45      & \textbf{67} & \textbf{1.181} & 1.259 & 1.034  & \textbf{0.913}\\
$C2/m$ (No. 12)    & Monoclinic      &    55     & \textbf{72}     &  53     & \textbf{67}  & \textbf{1.051} & 1.227 & \textbf{1.233} & 1.660 \\
$Pnma$ (No. 62)    & Orthorhombic    &    77     & \textbf{83}     & \textbf{76}     & 66 & \textbf{0.594}   & 1.092 & \textbf{0.313} & 1.633 \\
$I4/mmm$ (No. 139)   & Tetragonal      &   \textbf{91}      & 81     & \textbf{88}    & 63 & \textbf{0.140}  & 0.675 & \textbf{0.240} & 1.100\\
$R\bar{3}m$ (No. 166)   & Trigonal        &    \textbf{83}   & 74     &  \textbf{80}    &   71  & \textbf{0.294}  & 0.919 & \textbf{0.352} & 2.489\\
$P6_{3}/mmc$ (No. 194)   & Hexagonal       &   \textbf{97}     & 77     &  \textbf{96}    & 60 & \textbf{0.086}  & 0.545 & \textbf{0.324} & 4.100\\
$Fm\bar{3}m$ (No. 225)   & Cubic           &   \textbf{98}      & 96     & \textbf{95}     & 92 & \textbf{0.012}  & 0.033 & \textbf{0.214} & 0.483\\
\bottomrule
\label{table:700-stat}
\end{tabular}
\begin{tablenotes}
     \item [1] Calculated on the converged structures.
\end{tablenotes}
\end{threeparttable}
}
\end{table*}

Figs.~\ref{fig:initialization}a--g shows the average energy difference to the energy of final structures versus DFT relaxation steps. We neglected the structures whose energy changes and energy change intervals per step during relaxations exceeded 10 eV/atom to eliminate the impact of erroneous steps. One sees that \texttt{CrystalFormer}  samples generally reach lower energies in fewer relaxation steps. This is especially true for space groups with higher symmetries. The ability to initialize diverse and high-quality crystal structures enables one to discover more stable materials faster. Fig.~\ref{fig:initialization}h shows the histogram of energy above the convex hull constructed by the Materials Project database. The dashed line denotes the criterion  $E_\mathrm{hull} < 0.1$ eV/atom~\cite{sun2016thermodynamic} for selecting stable materials. Among these candidates, we found 34 and 12 relaxed structures with \texttt{CrystalFormer} and PyXtal initializations that are not contained in the MP-20 dataset. We summarize them in Table~\ref{table:samples} and Table~\ref{table:pyxtal_samples} of appendix~\ref{sec:initialization-app}. 

Table~\ref{table:700-stat} lists detailed statistics of structure-relaxed samples in seven representative space groups. Overall, we find that the \texttt{CrystalFormer} generated structures are of higher quality, especially for those space groups with higher symmetry. This observation is supported by the fact that the DFT relaxation often retains the space group symmetry.
The root mean squared displacement (RMSD)~\cite{zeni2023mattergen} computed for these converged structures demonstrate \texttt{CrystalFormer}'s superior performance over PyXtal across all 7 space groups.
The average energy above the convex hull also confirms the samples generated by \texttt{CrystalFormer} are indeed much closer to the DFT local minimum than PyXtal initialization. 
\texttt{CrystalFormer} attains superior performance in the high-symmetry space groups compared to the  RMSD of 0.11 $\text{\AA}$ reported in~\cite{zeni2023mattergen} for the MatterGen model trained on the MP-20 dataset with no control on the space-group symmetry.

\subsection{Structure-conditioned element substitution}
\label{sec:mutation}

Mutation of known crystals is a prominent approach to materials discovery. For example, one can employ a machine-learned force field to relax crystal structures~\cite{chen2022universal,merchant2023scaling,zhang2023dpa2, batatia2024foundation} after element substitutions. In the lens of generative modeling, the machine learning force field can be regarded as the energy-based model or Boltzmann machines. A potential drawback of exploring materials space with an energy-based model is the slow mixing or even ergodicity issue posed by the rough landscape of the potential energy surface. In this sense, element substitutions provide a variety of initial seeds, compensating for the limitation of energy-based exploration. Having an alternative measure of crystal likelihood other than the potential energy surface opens a way to employ the model likelihood as a guide for structure search. 



\begin{figure}[t]
    \centering
    \includegraphics[width=1\linewidth]{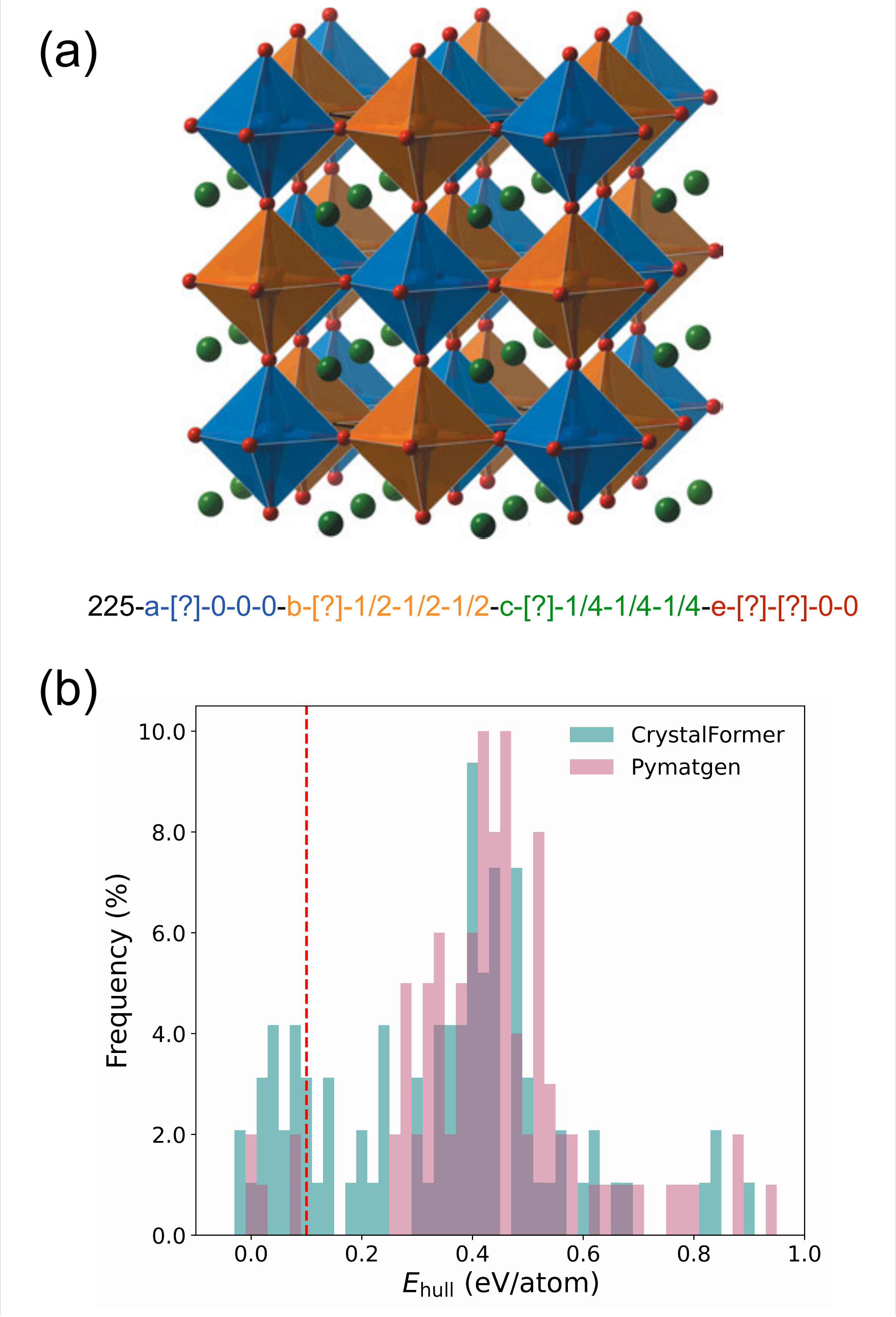}
    \caption{(a) Double perovskites crystal structure. The crystal string representation of double perovskites with blank spaces for chemical elements and the x-coordinate of the atom resides in the `e' position.  \texttt{CrystalFormer} generates crystals with double perovskite structures via sequence infilling. (b) The histograms of energy above the convex hull for the relaxed crystal structures. The dashed line indicates the criterion for selecting candidates for the stable materials.
    \label{fig:dp225}
    }
\end{figure}

Many crystal structures can be traced back to a few simple, highly symmetrical types. Numerous crystals share the same structural prototype but differ in composition, such as perovskite (ABX$_3$), spinel (AB$_2$X$_4$), fluorite (AX$_2$), and so on. Fig.~\ref{fig:dp225}a shows double perovskite crystal structures A$_2$BB$^\prime$X$_6$ which belong to the $Fm\bar{3}m$ (No. 225) space group.  
There are hundreds of known double perovskites with significant interests in their
semiconducting, ferroelectric, thermoelectric, and superconducting properties~\cite{vasala2015a2b}. Finding more stable materials with this structure prototype using brute force enumeration and high-throughput calculation is a computationally demanding task~\cite{wang2024landscape}. We will generate new double perovskites with \texttt{CrystalFormer}  and demonstrate its advantage of over standard element substitution methods.

Fig.~\ref{fig:dp225}a shows the string representation of double perovskites. To generate candidates of double perovskites, we use \texttt{CrystalFormer} to carry out string in-filling tasks. Since the autoregressive sampling of the atoms is insufficient to take into account non-causal information in the sequence, we employ MCMC to sweep through the sequence and update chemical species and fractional coordinates~\cite{miao2019cgmh}. The acceptance rate for these MCMC updates makes use of the marginalized probability for elements and fractional coordinates as the lattice parameter that appears at the end of the sequence can be integrated. Only after the MCMC sampling has been thermalized, we sample the lattice parameters autoregressive to account for the adjustment of the unit cell for given atoms and occupations. We use \texttt{CrystalFormer} to generate 100 candidates as the initial DFT relaxation. 

As a comparison, we also employ the \texttt{SubstitutionPredictorTransformation} function~\cite{hautier2011data} implemented in pymatgen~\cite{ong2013python} to perform element substitution for the crystals with double perovskite structures in the training dataset. The substitution probabilities come from data-mining of ICSD dataset~\cite{hautier2011data}. After the substitution, we use \texttt{DLSVolumePredictor}~\cite{CHU2018184} function of pymatgen to predict the volume of the structure. This lattice scaling scheme relies on data-mined bond lengths to predict the crystal volume of a given structure. To collect 100 candidates in the ionic substitution approach we have set the probability threshold of  \texttt{SubstitutionPredictorTransformation} to 0.01, which is smaller than the typical values adopted in Ref.~\cite{wang2021predicting}. 

The RMSD computed for the DFT-relaxed structures is 0.084 and 0.031 $\AA$ for \texttt{CrystalFormer} and ionic substitution~\cite{hautier2011data}, respectively. Moreover, Fig.~\ref{fig:dp225}b shows the histogram of energy above the convex hull of the Materials Project database. Overall, \texttt{CrystalFormer} and ionic substitution~\cite{hautier2011data} found 9 and 3 double perovskites with $E_\mathrm{hull}<0.1$eV/atom which are not contained in the MP-20 dataset, details in Appendix~\ref{sec:mutation-app}. The superior performance of \texttt{CrystalFormer}-guided MCMC is understandable since its likelihood takes into account the context of space group and atomic environment rather than marginal two-body correlation~\cite{hautier2011data} in ionic substitution. The ionic substitution approach also shows two limitations in practical applications. First, some of the ions in the compound can not be substituted as they are missing in the probability table. Second, the approach relies on the calculability of the elements' valence states which are not always well defined. 

As a final remark, although the discussion here focuses on generating crystals with given prototype structures, the generation of crystals with a given crystal lattice~\cite{okabe2024structural} is also feasible with \texttt{CrystalFormer}. This is because the crystal lattice can be straightforwardly expressed as constraints on the space group and occupied Wyckoff letters~\cite{regnault2022catalogue}.

\subsection{Plug-and-play materials design}
\label{sec:plug-and-play}

\begin{figure}[h]
    \centering
    \includegraphics[width=\linewidth]{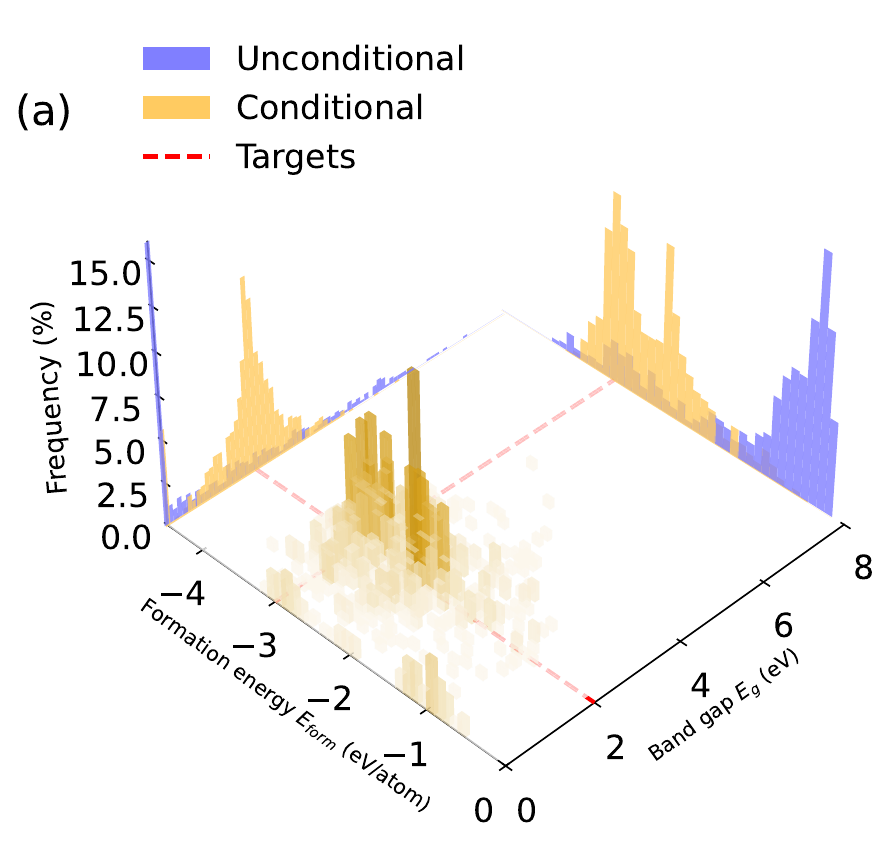}
    \includegraphics[width=\linewidth]{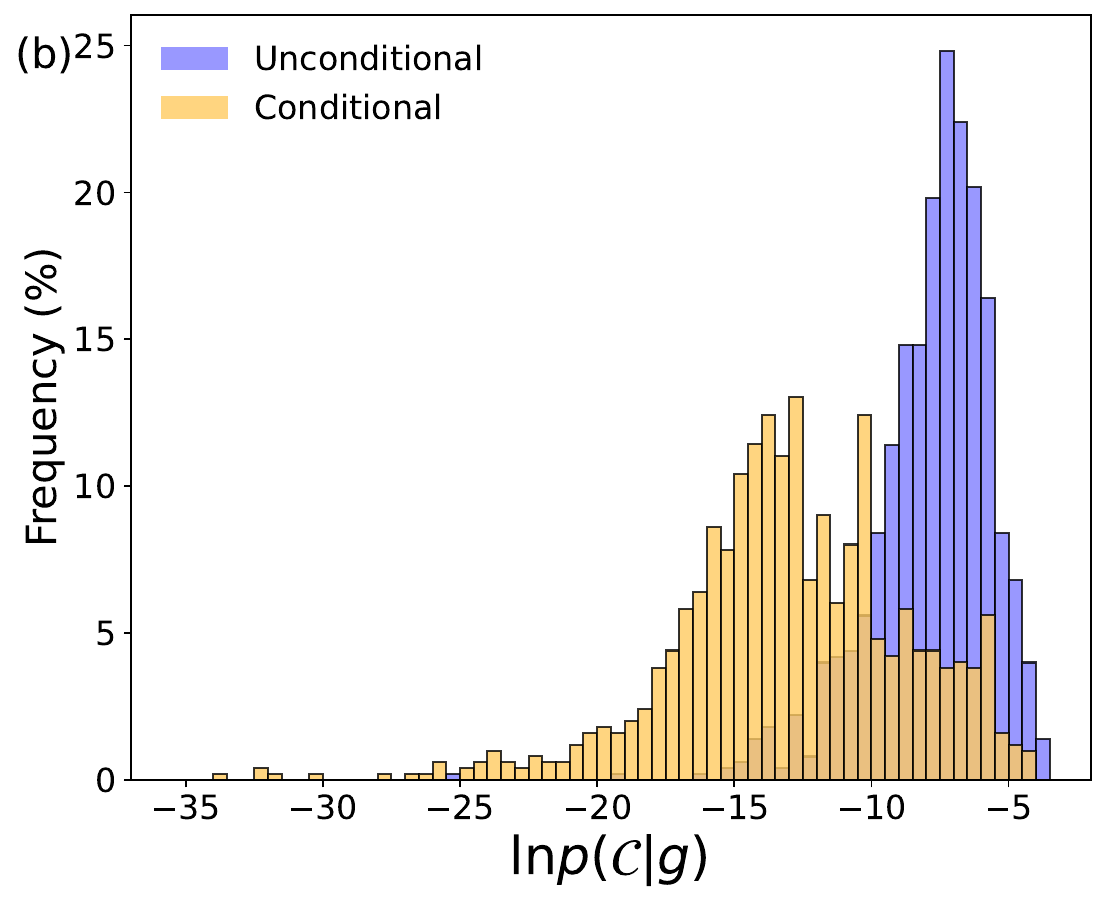}
    \caption{(a) The histogram of band gap and formation energy predicted by MEGNet models for crystal samples generated in the  $Fm\bar{3}m$ space group (No. 225). The dashed red lines in the plane indicate target values. The marginals on the side show the shift of the property distributions with respect to the unconditionally generated samples. Note that we scale the 3d histograms for better visualization. (b) The likelihoods of conditioned generated samples compared to the unconditional samples. 
    }
    \label{fig:plug-and-play}
\end{figure}

Finally, we demonstrate \texttt{CrystalFormer}'s ability to aid property-guided exploration of crystalline materials in a versatile and flexible manner. The trained \texttt{CrystalFormer} captures the space group conditioned crystal probability $ p(\boldsymbol{\mathcal{C}} |g )$, which we treat as a prior probability for stable crystals. By combining it with a crystal property prediction model that provides the forward likelihood probability $ p(y | \boldsymbol{\mathcal{C}} )$, one can carry out property-guided materials generation in a plug-and-play manner. According to Bayes' rule, the posterior for crystals given property $y$ reads 
\begin{equation}
p(\boldsymbol{\mathcal{C}} |g, y) \propto p(y | \boldsymbol{\mathcal{C}})  p(\boldsymbol{\mathcal{C}} |g ). 
\label{eq:bayes}
\end{equation}
By sampling from this posterior distribution, one can generate crystal samples with property guidance. Since the posterior probability \Eq{eq:bayes} typically does not process autoregressive property with respect to $\boldsymbol{\mathcal{C}}$, we carry out MCMC sampling to sample from the posterior distribution~\cite{verkuil2022language}. The plug-and-play feature makes designing crystalline materials in this way particularly appealing because it is possible to apply multiple conditions by simply adding log-likelihoods from multiple predictors. The framework applies to the inverse problem of solving cyrystal structures based on experimently observed diffraction spectra equally well~\cite{meredig2013hybrid,parackal2023identifying}, where the goal is to simultaneously optimizing the matching probability to experimental observation and stability of the crystal.

Any property prediction model can be used in conjunction with \texttt{CrystalFormer} for property-guided material generation. 
We  utilize two pre-trained MEGNet \cite{chen2019graph,chen2021learning} models to predict the band gap and formation energy, using the output of these two property prediction models as the forward probability $p(y | \boldsymbol{\mathcal{C}})$ of  \Eq{eq:bayes}. More details are in appendix~\ref{sec:plug-and-play-app}.

Fig.~\ref{fig:plug-and-play}a demonstrates the controlled generation of materials with target band gap at $E_\mathrm{g} = 2 ~\mathrm{eV}$ and the formation of energy $E_\mathrm{form} = -3 ~ \mathrm{eV/atom}$ crystals~\cite{franceschetti1999inverse, cheng2023global}. The conditional probability \Eq{eq:bayes} contains both the model likelihood and the property regression MAE. Therefore, the generated samples will strike a balance between the two. To draw samples from the conditional probability distribution, we randomly generate a batch of 1000 crystal samples and sweep through the crystal sequence to update the atom species, fractional coordinates, and lattice parameters~\footnote{For simplicity of the Wyckoff sequence are kept unchanged in the MCMC sampling.}. Achieving the desired properties via Monte Carlo update of chemical species can be regarded as a systematic data-informed way of carrying out cation-transmutation for materials inverse design~\cite{zhao2017design}.  After reaching equilibrium, the histogram of band gap and formation energy is centered around the target values, which is shifted significantly away from the value of unconditionally generated samples. On the other hand, the likelihood shown in Fig.~\ref{fig:plug-and-play}b indicates these conditional-generated crystals are not typical samples with respect to the unconditional distribution. Nevertheless, they are still probable samples according to the crystal prior given by the \texttt{CrystalFormer}. Note that controlling the distribution of formation energy can significantly impact the distribution of the energy above the convex hull due to the correlation between these two energies.
To assess the stability of the conditionally generated samples, we first use M3GNet \cite{chen2022universal} to filter out the unstable materials, followed by DFT verification
 on the remaining subset. The conditionally generated samples with  $E_\mathrm{hull}<0.1$ eV/atom  that are not included in the MP-20 dataset are listed in Table \ref{table:plug_and_play_samples} in appendix~\ref{sec:plug-and-play-app}. 
We observed that some materials meet the requirements of property predictors, but not the DFT calculations. This is due to
due to the errors in the property prediction models, which could be further improved by employing a more robust crystal property prediction models. Given a myriad of materials property prediction models developed over the years and the inconvenience of re-training or fine-tuning the foundational generative model~\cite{zeni2023mattergen}, we envision the plug-and-play generation approach demonstrated here to be a scalable way for materials design. We have exposed an interface of \texttt{CrystalFormer} in our code repository for users to plug in arbitrary conditioners for guided materials generation.

\section{Discussion and Conclusion}
\subsection{Related works}
\label{sec:related}

Crystal generative models have been explored using variational autoencoder~\cite{ren2022invertible,zhu2023wycryst}, generative adversarial networks~\cite{zhao2021high,luo2023towards}, normalizing flows~\cite{ahmad2022free, wirnsberger2022normalizing, kohler2023rigid, miller2024flowmm}, 
diffusion models~\cite{xie2021crystal, luo2023towards, jiao2023crystal, jiao2024space, zheng2023predicting, yang2023scalable, zeni2023mattergen,luo2024deep,ye2024concdvae}, GFlowNet~\cite{ai4science2023crystal, nguyen2023hierarchical}, and autoregressive models~\cite{gebauer2018generating, gebauer2019symmetry, gebauer2022inverse, xiao2023invertible, flam2023language, antunes2023crystal,gruver2024finetuned}. In these autoregressive models, one either uses atomistic features~\cite{gebauer2018generating, gebauer2019symmetry, gebauer2022inverse,xiao2023invertible}  or uses pure text descriptions~\cite{flam2023language, antunes2023crystal, gruver2024finetuned}. Nevertheless, with the introduction of specialized tokens for crystals, the boundary between the two is blurred. 

The \texttt{CrystalFormer} is most closely related to the autoregressive generative model originally designed for molecules~\cite{gebauer2018generating, gebauer2019symmetry, gebauer2022inverse}. However, instead of predicting the relative distances of atoms, we predict the Wyckoff positions of symmetry-inequivalent atoms in the unit cell. Having the luxury of the space group symmetry for crystals provides strong hints on where to put the atoms in the unit cell and greatly simplifies the design around spatial symmetries. On the other hand, compared to Ref.~\cite{antunes2023crystal} which treats text descriptions of crystals using autoregressive nature language model, \texttt{CrystalFormer} speaks native crystal language: it deals with a more concise and essential atomistic representation of crystals, which leads to a smaller model size and faster sampling speed. Fast generation speed is not only a welcoming feature but also will be crucial for further exploration of materials space based on combinations of probabilistic generation and post-selection, Monte Carlo sampling, backtracking, and searching techniques~\cite{yao2023tree}. More importantly, by baking in the space group symmetry in the model rather than learning them as statistical correlation from texts~\cite{flam2023language,gruver2024finetuned}, \texttt{CrystalFormer} guarantees space group constraints and cherishes the precious data and computing time. In this sense, the present work employs rigorous mathematical (as opposed to vague natural) language to incorporate the symmetry principle in the generative modeling of crystals. 

As a side remark, the Wyckoff position features have been used in machine learning models for materials property prediction~\cite{jain2018atomic, goodall2022rapid}. Incorporating space group information in the encoder-only transformer models may also enhance their property prediction performance~\cite{yan2022periodic, taniai2024crystalformer, xu2024predicting} as suggested by Ref.~\cite{rubungo2023llmprop}.

\subsection{Outlook}
\label{sec:outlook}

Precisely controlling the space group in the generative model of crystalline materials not only greatly simplifies the task but also is a highly desired feature for materials discovery and design. \texttt{CrystalFormer} integrates exact symmetry principles from math and empirical chemical intuitions from data into one unified framework. Probabilistic generative modeling of crystalline materials using \texttt{CrystalFormer} opens the way to many future innovations in materials design and discovery. 

Note that the MP-20 dataset has by no means exhausted all available crystalline material~\cite{antunes2023crystal,zeni2023mattergen}. An obvious future direction is to scale up the model as well as the training dataset, especially curating a dataset with better coverage of space groups. A later version of \texttt{CrystalFormer} is trained on curated Alex-20 dataset~\cite{schmidt2024improving, alex20} has shown significantly improved performance~\cite{cao2025crystalformer}. In particular, extending the dataset to include both inorganic and organic crystals~\cite{groom2016cambridge} may be beneficial as it improves the data coverage of low symmetric space groups. The transformer-based generative model is ready to be scaled up to work with much larger and more diverse training data, in the same fashion as large language models~\cite{brown2020language}. Given similar model architectures, the idea of generative pretraining of a foundational model for material generation is appealing. When scaling up the model it will be interesting to note the possible appearance of neural scaling law~\cite{kaplan2020scaling} as it has also been showing up in other contexts of atomistic modeling~\cite{frey2023neural}. 

The model architecture and sampling strategy are both open to further refinement to better serve the purpose of material discovery. First of all, to better facilitate data efficiency learning and structure phase transitions-related applications, it will be useful to further exploit the Euclidean normalizer~\cite{muller2013symmetry} and group-subgroup relation~\cite{hatch1989isotropy} in the model architecture or training procedure.  Second, it is worth exploring using \texttt{CrystalFormer} as the base distribution in the flow model and employs symmetry-persevering transportation to further adjust the atoms coordinates and unit cells~\cite{ingraham2023illuminating, jiao2024space}, which mimics a symmetry-constrained relaxation process~\cite{cox2022symmetric}. Lastly, it may be worth employing more advanced constrained and guided sequence generation methods \cite{dathathri2019plug,zhang2020language,qin2022cold,lew2023sequential} for more flexible control on the elements, structure, or stoichiometry of generated materials.

Conditioned materials generation depending on  properties~\cite{gebauer2022inverse,zeni2023mattergen,luo2024deep,ye2024concdvae} and experimental measurements~\cite{lai2024end} are highly desired features of materials generative model. Although it is straightforward to extend \texttt{CrystalFormer} (e.g. extend the space group embedding or employ the encoder-decoder transformer architecture~\cite{vaswani2017attention}) to incorporate these conditions, we are particularly excited about the plug-and-play routine demonstrated in Sec.~\ref{sec:plug-and-play}. Along this line, we envision an ecosystem~\cite{raymond1999cathedral} where the foundational generative model for $p(\boldsymbol{\mathcal{C}}|g)$ and more specialized discriminative models for materials properties $p(y|\boldsymbol{\mathcal{C}})$ are developed separately but brought together via the Bayes rule. 



\section*{Conflict of Interest}
The authors declare that they have no conflict of interest.

\begin{acknowledgments}
We thank Han Wang, Lin Yao, Linfeng Zhang, Chen Fang, Yanchao Wang, Zhenyu Wang, Qi Yang, Shigang Ou, Xinyang Dong, Wenbing Huang, Quansheng Wu, Wanjian Yin, Xi Dai, Shuang Jia, Hangtian Zhu, Jiangang Guo, and Hongjian Zhao for useful discussions.
This project is supported by the National Natural Science Foundation of China (T2225018, 92270107, 12188101, T2121001, and 12034009), National Key Projects for Research and Development of China (2021YFA1400400), and the Strategic Priority Research Program of Chinese Academy of Sciences (XDB0500000 and XDB30000000). 

\section*{Author contributions}
Zhendong Cao and Lei Wang wrote the code and trained the model.
Xiaoshan Luo performed the numerical calculations.
Jian Lv and Lei Wang designed the research and supervised the project.
All authors contributed to the data analysis, discussion of the results, and preparation of the manuscript.

\section*{Data availability}
We have released the codes and trained model at \href{https://github.com/deepmodeling/CrystalFormer}{https://github.com/deepmodeling/CrystalFormer}.

\end{acknowledgments}

\bibliography{refs}

\clearpage
\newpage

\onecolumngrid

\maketitle
\setcounter{algorithm}{0}
\renewcommand{\thealgorithm}{\arabic{algorithm}}
\setcounter{table}{0}
\renewcommand{\thetable}{S\arabic{table}}
\setcounter{figure}{0}
\renewcommand{\thefigure}{S\arabic{figure}}
\setcounter{equation}{0}
\renewcommand{\theequation}{S\arabic{equation}}

\appendix

\section{More details of \texttt{CrystalFormer}}

To recap, the space group information plays a key role in the architecture, training, and sampling of \texttt{CrystalFormer}. First of all, the vector embedding of space group number $g$ controls all subsequent outputs of the transformer corresponding to the Wyckoff letters, chemical species, fractional coordinates, and lattice parameters. Second, the information of the space group and Wyckoff letter are used to select active components in the fractional coordinates and lattice parameters in the loss function during training. Lastly, the space group determines the concrete meaning of Wyckoff letters in terms of multiplicities and fractional coordinates, which are used to place the right number of atoms precisely in the unit cell during sampling. 

\subsection{Model architectures}
Algorithm~\ref{alg:model} summarized the model architecture of \texttt{CrystalFormer}. Training the model for 3,800 epochs with the hyperparameters shown in Table~\ref{table:hyperparameters} takes about 13 hours on a single A100 GPU. 
 
\begin{algorithm}[H]
\caption{The \texttt{CrystalFormer} architecture}
\begin{algorithmic}[1] 
\Require Space group number $g$, Wyckoff letters  $\boldsymbol{W}=[w_i]$,  multiplicity of Wyckoff positions $\boldsymbol{M}=[m_i]$, chemical elements $\boldsymbol{A}=[a_i]$, fractional coordinates $\boldsymbol{X}=[(x_i, y_i, z_i)]$ of each atom in the unit cell.
\Ensure Parameters for the conditional probability of Wyckoff letters $\boldsymbol{\omega}_i$, chemical element $\boldsymbol{\alpha}_i$, and fractional coordinates $\boldsymbol{\chi}_i, \boldsymbol{\upsilon}_i, \boldsymbol{\zeta}_i $ of atoms and the lattice parameters  $\boldsymbol{\ell} $. 
\State $\boldsymbol{\omega}_1 = \texttt{Net}(g) $  \Comment the logit of the first Wyckoff position is implemented as a standalone neural network. 
\State{ \href{https://github.com/deepmodeling/CrystalFormer/blob/d4ba4841a57c80a2b9fa9ee38fa8a11eff9ab5a2/crystalformer/src/transformer.py#L78-L105}{\textcolor{blue}{\# prepare input features} }}
\State  $\boldsymbol{h}_{\boldsymbol{W}} = [\texttt{Embed}(g), \texttt{Embed}(w_i),  m_i]$.  
\State  $\boldsymbol{h}_{\boldsymbol{A}} = [\texttt{Embed}(g), \texttt{Embed}(a_i) ]$.  
\State  $\boldsymbol{h}_{\boldsymbol{X}} = [\texttt{Embed}(g), \cos(2\pi x_i ), \sin (2\pi x_i ), \ldots,\cos(2\pi x_i N_f), \sin (2\pi x_i N_f ) ]  $                
\State  $\boldsymbol{h}_{\boldsymbol{Y}} = \ldots $ 
\State  $\boldsymbol{h}_{\boldsymbol{Z}} = \ldots $ 
\State{\href{https://github.com/deepmodeling/CrystalFormer/blob/d4ba4841a57c80a2b9fa9ee38fa8a11eff9ab5a2/crystalformer/src/transformer.py#L107-L151}{\textcolor{blue}{\# concatenate along particle dimension}}}
\State $\boldsymbol{h}= \mathtt{Concatenate}(\boldsymbol{h}_{\boldsymbol{W}}, \boldsymbol{h}_{\boldsymbol{A}}, \boldsymbol{h}_{\boldsymbol{X}}, \boldsymbol{h}_{\boldsymbol{Y}}, \boldsymbol{h}_{\boldsymbol{Z}} ) $ 
\State Project $\boldsymbol{h}$ feature size to $d_\mathrm{model}$ and add position embedding  
\State $\boldsymbol{h} = \texttt{MaskedTransformer}(\boldsymbol{h})$ 
\State Project $\boldsymbol{h}$ feature size to desired dimensions
\State{\href{https://github.com/deepmodeling/CrystalFormer/blob/d4ba4841a57c80a2b9fa9ee38fa8a11eff9ab5a2/crystalformer/src/transformer.py#L153-L222}{\textcolor{blue}{\# split along particle dimension}}}
\State $\boldsymbol{\omega}_i, \boldsymbol{\alpha}_i , \boldsymbol{\chi}_i, \boldsymbol{\upsilon}_i, \boldsymbol{\zeta}_i,  \boldsymbol{\ell} 
=\texttt{Split}(\boldsymbol{h} )$   
\State Mask $\boldsymbol{\omega}_i$ to ensure the Wyckoff letters are valid for the given space group $g$ and appear in alphabetical order.
\State \Return $[\boldsymbol{\omega}_1, \boldsymbol{\alpha}_1,  \boldsymbol{\chi}_1,  \boldsymbol{\upsilon}_1, \boldsymbol{\zeta}_1, \boldsymbol{\omega}_2,  \boldsymbol{\alpha}_2, \boldsymbol{\chi}_2, ... , \boldsymbol{\ell}] $
\end{algorithmic}
\label{alg:model}
\end{algorithm}

\newpage

\begin{table*}[h!]
\caption{A table of hyperparameters used in this work.}
\begin{center}
\renewcommand{\arraystretch}{1.2}
\setlength{\tabcolsep}{24pt} 
\begin{tabular}{lcl}
\toprule
\textbf{Hyperparameters} & \textbf{Value} & \textbf{Remarks} \\
\hline
The length of atom sequence including the padding atoms & 21 \\
Number of chemical species  &  119  & `H' to `Og', plus padding atom \\
Number of possible Wyckoff letters & 28 & `a-z'+`A', plus padding atom  \\
Number of modes in von-Mises mixture distribution $K_{x}$ & 16 \\
Number of modes in lattice Gaussian mixture distribution $K_{l}$ & 4  \\
Hidden layer dimension for the composite type of the first atom & 256 \\
Transformer number of layers & 16\\
Transformer number of heads & 16 \\
Transformer key size & 64 \\
Transformer model size  $d_\mathrm{model}$ & 32 \\
Embedding dimension of discrete input  & 32 \\
Number of Fourier frequency $N_{f}$ & 5 \\
Learning rate & 0.0001 \\
Learning rate decay  & 0.0 \\
Weight decay & 0.0 \\
Clip grad & 1.0 \\
Batch Size & 100 \\
Optimizer & Adam \\
Dropout rate & 0.5 \\
\hline
Total number of parameters:  4840295 \\
\bottomrule
\end{tabular}
\end{center}
\label{table:hyperparameters}
\end{table*}

\newpage

\subsection{Sampling algorithm}

Algorithm~\ref{alg:sample} summarizes the autoregressive sampling method of \texttt{CrystalFormer}. It takes 520 seconds to generate a batch size 13,000 crystal samples on a single A100 GPU, which translates to a generation speed of 40 milliseconds per sample.

\begin{algorithm}[H]
\caption{Autoregressvie sampling of \texttt{CrystalFormer}}
\begin{algorithmic}[1]
\Require space group number $g$,  a list of chemical elements \texttt{element\_list}, maximum number of Wyckoff positions $n$, sampling temperature $T$
\Ensure Wyckoff letters $\boldsymbol{W}$, chemical species $\boldsymbol{A}$,  fractional coordinates $\boldsymbol{X}$ of atoms, and lattice parameters $\boldsymbol{L}$ of the unit cell. 
\State Initialize $\boldsymbol{W}=[\emptyset]$, $\boldsymbol{A}=[\emptyset]$,  $\boldsymbol{X}=[\emptyset]$ 
\For{$i = 1 \ldots, n$}
    \State{\href{https://github.com/deepmodeling/CrystalFormer/blob/d4ba4841a57c80a2b9fa9ee38fa8a11eff9ab5a2/crystalformer/src/sample.py#L68-L76}{\textcolor{blue}{\# sample Wyckoff letter $w$}}}
        \State Get the last $ \boldsymbol{\omega}$ from  $\texttt{CrystalFormer}(g, \boldsymbol{W}, \boldsymbol{A}, \boldsymbol{X}  )$ 
        \State $w  \sim \mathrm{Categorical}(\boldsymbol{\omega})^{1/T}$ 
        \State $\boldsymbol{W}[i] =  w $ 
    \State{\href{https://github.com/deepmodeling/CrystalFormer/blob/d4ba4841a57c80a2b9fa9ee38fa8a11eff9ab5a2/crystalformer/src/sample.py#L78-L93}{\textcolor{blue}{\# sample atom species $a$}}}
     \State Get the last $ \boldsymbol{\alpha}$ from  $\texttt{CrystalFormer}(g, \boldsymbol{W}, \boldsymbol{A}, \boldsymbol{X}  )$ 
        \State Mask the logits in $\boldsymbol{\alpha}$ according to \texttt{element\_list}
        \State $a  \sim \mathrm{Categorical}(\boldsymbol{\alpha})^{1/T}$ 
        \State $\boldsymbol{A}[i] =  a $
        \State{\href{https://github.com/deepmodeling/CrystalFormer/blob/d4ba4841a57c80a2b9fa9ee38fa8a11eff9ab5a2/crystalformer/src/sample.py#L98-L109}{\textcolor{blue}{\# sample fractional coordinate $x$}}}
        \State Get the last $\boldsymbol{\chi}$ from $ \texttt{CrystalFormer}(g, \boldsymbol{W},  \boldsymbol{A}, \boldsymbol{X})$ 
        \State $ {x} \sim \mathrm{vonMisesMix}(\boldsymbol{\chi})^{1/T}$
        \State Project ${x}$ to Wyckoff positions according to the Wyckoff letter $w$
        \State update $\boldsymbol{X}$ with $x$
  \State{\href{https://github.com/deepmodeling/CrystalFormer/blob/d4ba4841a57c80a2b9fa9ee38fa8a11eff9ab5a2/crystalformer/src/sample.py#L111-L122}{\textcolor{blue}{\# sample fractional coordinate $y$}}}
      \State Get the last $\boldsymbol{\upsilon}$ from $ \texttt{CrystalFormer}(g, \boldsymbol{W},  \boldsymbol{A}, \boldsymbol{X})$ 
      \State ... 
      \State update $\boldsymbol{X}$  with $y$
 \State{\href{https://github.com/deepmodeling/CrystalFormer/blob/d4ba4841a57c80a2b9fa9ee38fa8a11eff9ab5a2/crystalformer/src/sample.py#L124-L135}{\textcolor{blue}{\# sample fractional coordinate $z$}}}
    \State Get the last $\boldsymbol{\zeta}$ from $ \texttt{CrystalFormer}(g, \boldsymbol{W},  \boldsymbol{A}, \boldsymbol{X})$ 
    \State ... 
    \State update $\boldsymbol{X}$  with $z$

\EndFor
\State{\href{https://github.com/deepmodeling/CrystalFormer/blob/d4ba4841a57c80a2b9fa9ee38fa8a11eff9ab5a2/crystalformer/src/sample.py#L153-L172}{\textcolor{blue}{\# sample $\boldsymbol{L}$}}}
\State Get $ \boldsymbol{\ell} $ from $\texttt{CrystalFormer}(g, \boldsymbol{W}, \boldsymbol{A}, \boldsymbol{X})$
\State $\boldsymbol{L} \sim \mathrm{GaussianMix}(\boldsymbol{\ell})^{1/T}$
\State Symmetrize $\boldsymbol{L}$ according to space group $g$
\State \Return  $\boldsymbol{W}$, $\boldsymbol{A}$, $\boldsymbol{X}$, $\boldsymbol{L}$
\end{algorithmic}
\label{alg:sample}
\end{algorithm}

\newpage

Algorithm~\ref{alg:mcmc} summarizes the Markov chain Monte Carlo sampling of \texttt{CrystalFormer}. It is used in the structure constrained generation of crystals in Sec.~\ref{sec:mutation}. The property-guided materials design discussed in Sec.~\ref{sec:plug-and-play} employs a similar sampling strategy for the posterior probability distribution \Eq{eq:bayes}. 

\begin{algorithm}[H]
\caption{Markov chain Monte Carlo sampling of \texttt{CrystalFormer}}
\begin{algorithmic}[1]
\Require Space group number $g$, Wyckoff letters $\boldsymbol{W}$, chemical species $\boldsymbol{A}$,  fractional coordinates $\boldsymbol{X}$ of atoms, and lattice parameters $\boldsymbol{L}$ of the unit cell, length $n$ of the atom sequence, a list of chemical elements \texttt{element\_list}, step size $\epsilon$,
\Ensure Wyckoff letters $\boldsymbol{W}$, chemical species $\boldsymbol{A}$,  fractional coordinates $\boldsymbol{X}$ of atoms, and lattice parameters $\boldsymbol{L}$ of the unit cell.

\State $\boldsymbol{\mathcal{C}} = (\boldsymbol{W}, \boldsymbol{A}, \boldsymbol{X}, \boldsymbol{L})$

\For{$i = 1 \ldots, $ steps}
\For{$j = 1 \ldots, n$}
    \State{\href{https://github.com/deepmodeling/CrystalFormer/blob/d4ba4841a57c80a2b9fa9ee38fa8a11eff9ab5a2/crystalformer/src/mcmc.py#L53-L57}{\textcolor{blue}{\# update element $a$}}}
        \State $a' \sim $ \texttt{element\_list}
    \State{\href{https://github.com/deepmodeling/CrystalFormer/blob/d4ba4841a57c80a2b9fa9ee38fa8a11eff9ab5a2/crystalformer/src/mcmc.py#L59-L64}{\textcolor{blue}{\# update coordinate $\boldsymbol{x} $}}}
        \State  $\Delta \boldsymbol{x} \sim$ vonMises
        \State Mask the $\boldsymbol{x}$ according to the space group $g$ and Wyckoff letter $w$
    \State{\href{https://github.com/deepmodeling/CrystalFormer/blob/d4ba4841a57c80a2b9fa9ee38fa8a11eff9ab5a2/crystalformer/src/mcmc.py#L66-L76}{\textcolor{blue}{\# update $\boldsymbol{\mathcal{C}}$ }}}
        \State Propose an update $\boldsymbol{\mathcal{C}} \rightarrow \boldsymbol{\mathcal{C'}}$ with $a \rightarrow a'$ and $\boldsymbol{x} \rightarrow \boldsymbol{x}+ \epsilon \Delta \boldsymbol{x}$ 
        \State Update according to the Metropolis acceptance probability  $\min \left[1, \frac{p(\boldsymbol{\mathcal{C'}}|g)}{p(\boldsymbol{\mathcal{C}}|g)}\right]$

\EndFor
\EndFor
\State{\href{https://github.com/deepmodeling/CrystalFormer/blob/d4ba4841a57c80a2b9fa9ee38fa8a11eff9ab5a2/crystalformer/src/sample.py#L183-L221}{\textcolor{blue}{\# update lattice $\boldsymbol{L}$}}}
\State Get $ \boldsymbol{\ell} $ from $\texttt{CrystalFormer}(g, \boldsymbol{W}, \boldsymbol{A}, \boldsymbol{X})$
\State $\boldsymbol{L} \sim \mathrm{GaussianMix}(\boldsymbol{\ell})^{1/T}$
\State Symmetrize the $\boldsymbol{L}$ according to the space group $g$

\State \Return  $\boldsymbol{W}$, $\boldsymbol{A}$, $\boldsymbol{X}$, $\boldsymbol{L}$
\end{algorithmic}
\label{alg:mcmc}
\end{algorithm}

\newpage

\section{Validity and novelty of generated samples}
\label{sec:validity}

\begin{sidewaysfigure}
    \centering   
    \includegraphics[width=1\textwidth]{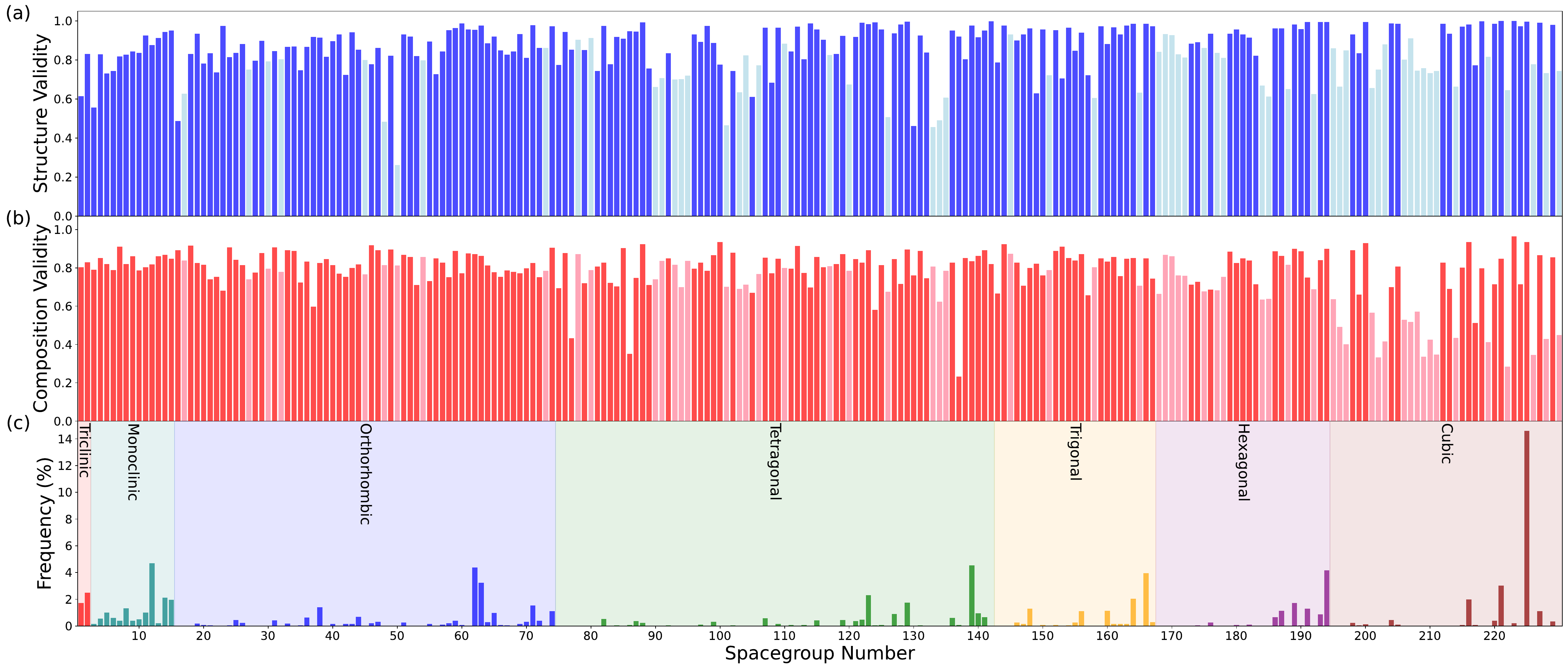}
    \caption{(a) Structure validity histogram, (b) Composition validity histogram (c) frequency of space groups in the training dataset. The histograms in light colors in (a) and (b) are for those space groups where there is no training data. A dataset of 1000 generated samples per space group was curated for evaluation.}
    \label{fig:valid_hist}
\end{sidewaysfigure}

Figure~\ref{fig:valid_hist} illustrates the structure and compositional validity of generated samples across all 230 space groups. Following the Ref.~\cite{court20203}, a structure meets the validity criteria if the shortest atomic distance exceeds $0.5~\AA$, a lenient standard. Composition validity requires charge neutrality as computed by SMACT \cite{davies2019smact}. This is, however, an overly stringent criterion since the composition validity of the training set is only around  
$90\%$ by this measure~\cite{xie2021crystal}. 

\begin{table}[h]
\begin{threeparttable}
\centering
\caption{Validity of generated crystal structure for representative space groups. Training samples count the number of samples in the training set.}
\renewcommand{\arraystretch}{1.2}
\begin{tabular}{llcccc}
\toprule
\multirow{2}{*}{\textbf{Space group}} & \multirow{2}{*}{\textbf{Crystal system}} & \multirow{2}{*}{\textbf{Training samples}} & \multicolumn{2}{c}{{\textbf{Validity ($\%$) $\uparrow$}}} & \textbf{Params}  \\
    &      &                &  Struc.      & Comp.               \\ 
\midrule
2     & Triclinic     & 676    &    83.10     &     83.0            \\
12    & Monoclinic    & 1273   &    87.70     &      81.80           \\
62    & Orthorhombic  & 1187   &    95.50     &      87.20           \\
139   & Tetragonal    & 1233   &    97.70     &      83.40           \\
166   & Trigonal      & 1076   &    98.50     &      85.0           \\
194   & Hexagonal     & 1129   &    99.40     &      89.90           \\
225   & Cubic         & 3960   &    99.60     &       93.50         \\
\midrule
1     & Triclinic     &   27136   &  91.40    &     80.20        \\
\midrule
\multicolumn{3}{l}{\textbf{Autoregressive models}} \\
\multicolumn{3}{l}{\quad 
PGSchNet~\cite{gebauer2019symmetry}}  & 99.65 &  75.96 & -\\
\multicolumn{3}{l}{\quad LM-CH (character-level tokenization)~\cite{flam2023language}}   & 84.81 & 83.55 & 1$\sim$100 M  \\ 
\multicolumn{3}{l}{\quad LM-AC (atom coordinate-level tokenization)~\cite{flam2023language} }   & 95.81 & 88.87 & 1$\sim$100 M \\ 
\multicolumn{3}{l}{\quad Crystal-text-LLM~\cite{gruver2024finetuned} }   & 96.5 & 86.3 & 70 B \\ 
\multicolumn{3}{l}{\textbf{Diffusion models} } \\
\multicolumn{3}{l}{\quad CDVAE~\cite{xie2021crystal} }  & 100.0 & 86.70 & 4.5 M \\
\multicolumn{3}{l}{\quad DiffCSP~\cite{jiao2023crystal} }  & 100.0 & 83.25 & 12.3 M \\
\multicolumn{3}{l}{\quad DiffCSP++~\cite{jiao2024space} }  & 99.94 & 85.12 &  12.3 M\\
\multicolumn{3}{l}{\quad UniMat-Large~\cite{yang2023scalable}}  & 97.2 & 89.4 & - \\
\bottomrule
\label{table:validity_table}
\end{tabular}
\end{threeparttable}
\end{table}

Table~\ref{table:validity_table} reports the validity of generated samples for selected space groups in each of the seven crystal systems. To ensure the numbers are representative, we chose the space group to be the one with the most training data for each crystal system. One sees that the model performs better for more symmetric space groups. 
This is a nice feature that complements existing crystal generative models, which mostly have difficulties in generating highly symmetric structures. As a reference, we also list the validity of the generated samples suppose one treats the crystals as if they are all in the $P1$ space group (No. 1) with only translational symmetry. One sees the structure validity scores of $P1$ space group improve compared to the one shown in Figure~\ref{fig:valid_hist} due to increased training samples. 

The second part of Table~\ref{table:validity_table} shows reference results in the literature for the same validity test. In principle, the performance of the present model should fall back to the language model approaches~\cite{flam2023language, gruver2024finetuned}. The remaining gap may be due to details such as including the header line in the crystallographic information files (CIF), specific sampling strategy of language models, or the additional post-selection of samples~\cite{gruver2024finetuned}.  In the table, the DiffCSP++~\cite{jiao2024space} is the only alternative model that exploits the space group symmetry in a rigorous manner similar to ours. However, the DiffCSP++ does not predict Wyckoff positions in the generation process. Instead, one needs to search for template structures in the training set for generation, which may limit its generality. Besides works listed in Table~\ref{table:validity_table} that reported validity scores in comparable settings, Ref.~\cite{antunes2023crystal} has conditioned the generation of CIF on the space group symbols in a language model setting. Ref.~\cite{zeni2023mattergen} considered space group conditioned crystal generation using a fine-tuned generative model with space group labels. Neither approach provides exact constraints on the space group, which could yield problematic structures for large systems and highly symmetric space groups. Ref.~\cite{zhu2023wycryst} considered generating symmetric crystals in their Wyckoff representation. However, the model does not consider fractional coordinates and lattice parameters, so it requires a subsequent computational search to completely determine the crystal structure. 


\begin{figure}[h]
    \centering
    \includegraphics[width=0.6\linewidth]{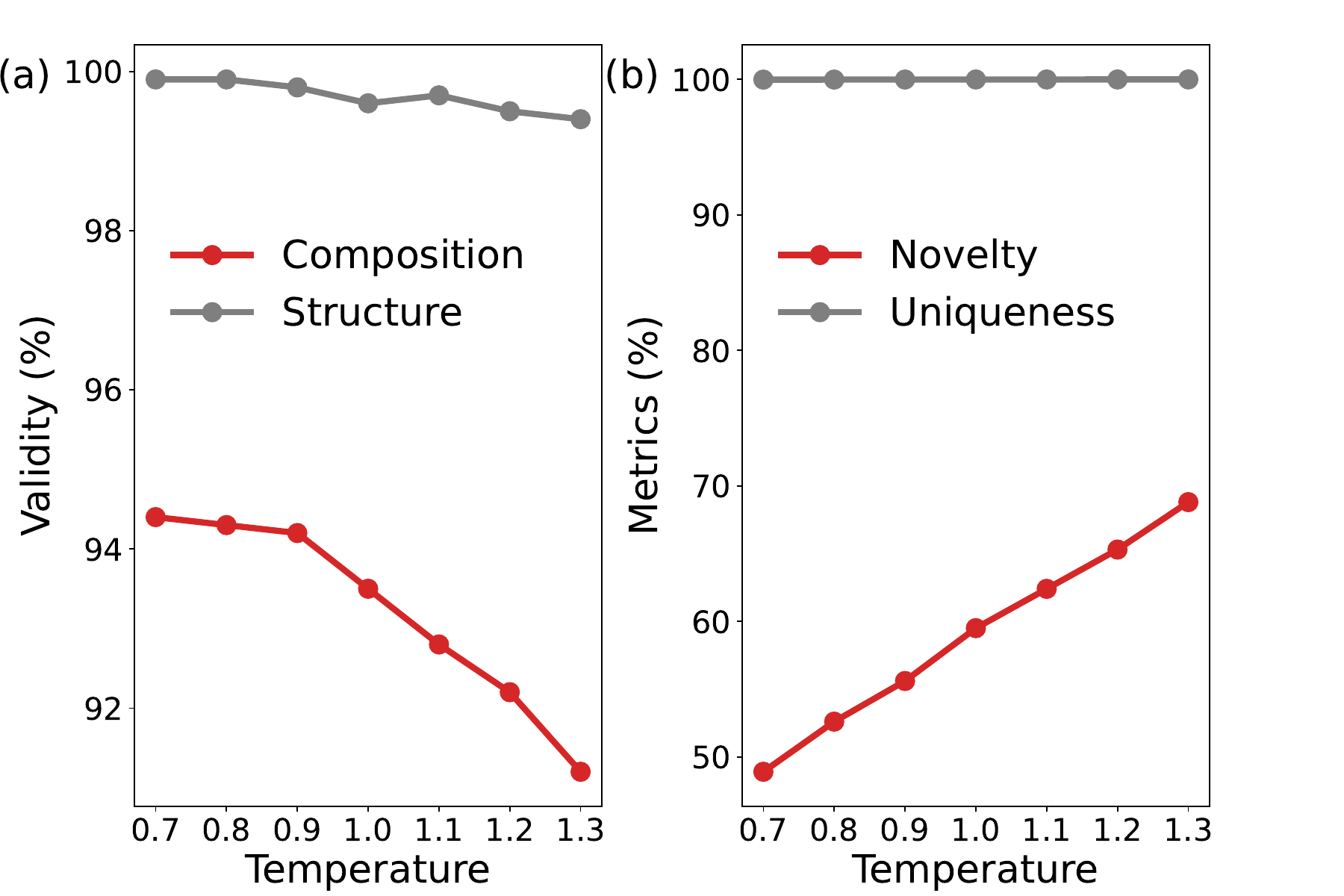}
    \caption{(a) Structure and composition validity,  (b) novelty and uniqueness of generated samples evaluated according to~\cite{xie2021crystal} for the $Fm\bar{3}m$ space group. 
    }
    \label{fig:validity}
\end{figure}

Figure~\ref{fig:validity}(a) shows the validity of generated samples as a function of sampling temperature. One clearly sees that reduced temperature $T<1$ increases the validity of samples at the cost of reducing the diversity~\cite{gruver2024finetuned}. Figure~\ref{fig:validity}(b) shows the novelty and uniqueness evaluated on $1000$ generated samples with temperature. Novelty quantifies the proportion of new structures in the generated samples that were unseen in the training dataset. Uniqueness represents the percentage of distinct, non-redundant structures among the generated samples~\cite{matbench_genmetrics}. One sees that across different temperatures the uniqueness remains high, indicating the model does not collapse to a mode that produces duplicated samples. On the other hand, the model produces close to $70\%$ novel material as temperature increases, which nicely demonstrates modal covering behavior of the maximum likelihood estimation training~\cite{Goodfellow-et-al-2016}. Having a model distribution broader than the span of the dataset is crucial for material discovery. 

\onecolumngrid
\clearpage
\clearpage
\twocolumngrid

\onecolumngrid

\section{Discovered crystal samples with symmetric structure initialization}
\label{sec:initialization-app}

Table~\ref{table:samples} and Table~\ref{table:pyxtal_samples} list novel samples with $E_\mathrm{hull} < 0.1$ eV/atom with \texttt{CrystalFormer} and PyXtal initializations respectively. Among them, Pd$_3$Pt, MgAgPd$_2$, TaNbRu$_2$ and Gd$_2$HgAu are both discovered by \texttt{CrystalFormer} and PyXtal initialization. These compounds share identical crystal structures in the two tables, with lattice constants differing by less than 0.02 $\AA$.

To relax samples and estimate the energy above the hull, the DFT calculations were performed with the Perdew-Burke-Ernzerhof (PBE) exchange-correlation functional~\cite{Perdew.PBE.1996} and all-electron projector-augmented wave method~\cite{Blochl.PAW.1994}, as implemented in the VASP code~\cite{Kresse.VASP.1996}. All parameters of the calculations including settings of PBE functional, Hubbard U corrections, and ferromagnetic initialization are chosen to be consistent with Materials Project by using of \texttt{\detokenize{MPRelaxSet}} function in pymatgen~\cite{ong2013python} A double relaxation strategy was employed. The maximum optimization ionic step and the maximum running time were constrained to 150 steps and 20 hours, respectively. All structures containing Yb element are ignored when calculating energy above the hull due to they are unavailable from the Materials Project at the time of writing.

\begin{figure*}[h]
    \centering
    \includegraphics[width=1\linewidth]{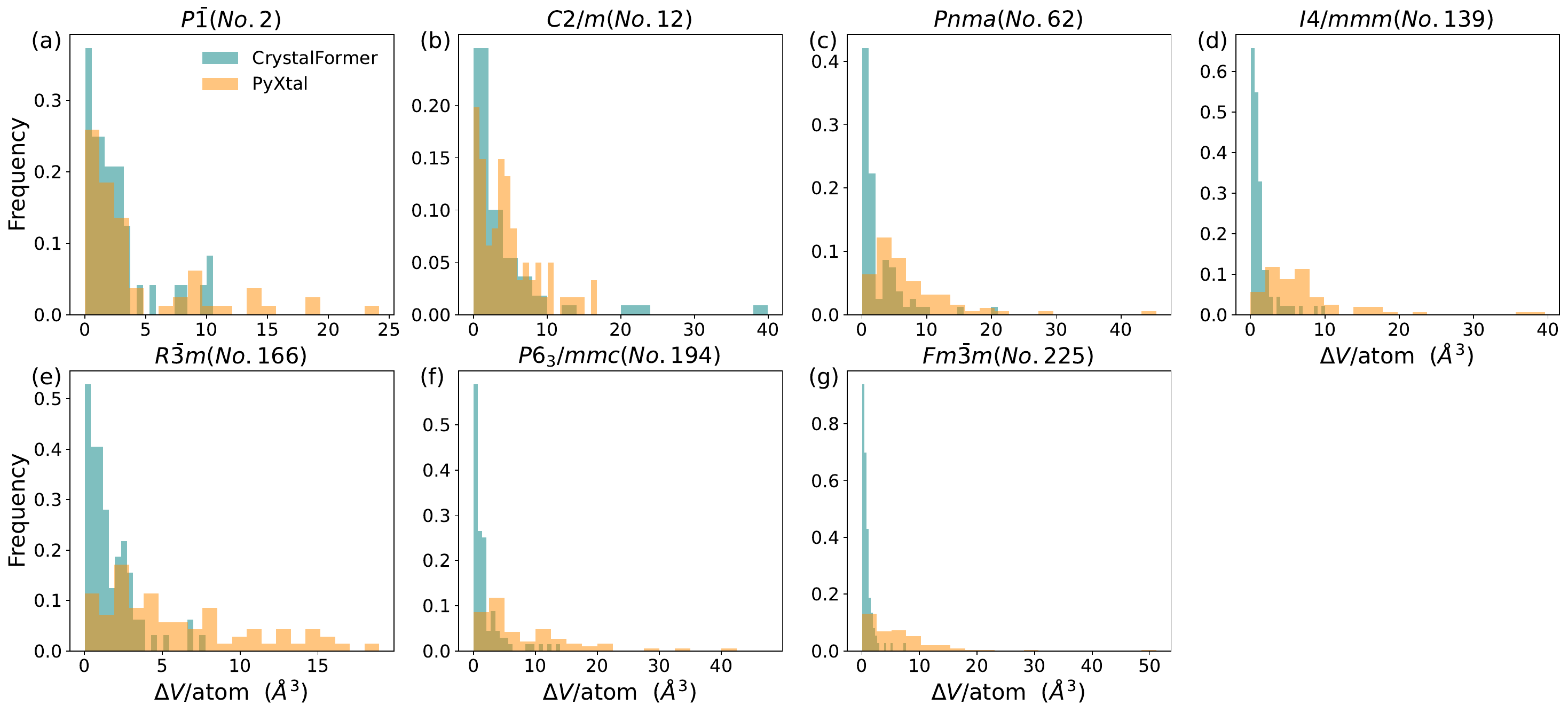}
    \caption{(a)-(g) The distribution of per-atom volume changes before and after relaxation in seven representative space groups.
    }
\end{figure*}

\begin{table}[t]
\begin{threeparttable}
\centering
\caption{Discovered crystalline materials by \texttt{CrystalFormer} with $E_\mathrm{hull} < 0.1$ eV/atom which are not in the MP-20 dataset.}
\renewcommand{\arraystretch}{1.2}
\begin{tabular}{lclr}
\toprule
\textbf{Formula} & \textbf{Space group} & \textbf{Wyckoff-Atom sequence}~\tnote{1}  &  \textbf{$E_\mathrm{hull}$ (eV/atom)} \\
\midrule
LuSiNi & 62 & c-Lu-c-Si-c-Ni & 0.0134 \\
Eu$_3$Mg & 62 & c-Eu-c-Mg-d-Eu & 0.0661 \\
EuAgAu & 62 & c-Eu-c-Ag-c-Au & 0.0111 \\
SmSbPd & 62 & c-Sm-c-Sb-c-Pd & 0.0368 \\
Nd$_3$Rh & 62 & c-Nd-c-Rh-d-Nd & 0.0673 \\
SrSnPd & 62 & c-Sr-c-Sn-c-Pd & -0.0531 \\
SrCdSn & 62 & c-Sr-c-Cd-c-Sn & 0.0863 \\
SrMgHg & 62 & c-Sr-c-Mg-c-Hg & 0.0517 \\
Eu$_2$Ru & 62 & c-Eu-c-Eu-c-Ru & 0.0890 \\
PrHgAu$_2$ & 62 & c-Pr-c-Hg-c-Au-c-Au & 0.0868 \\
Ce(SiPt)$_2$ & 139 & a-Ce-d-Si-e-Pt & 0.0648 \\
Dy(AlGa)$_2$ & 139 & a-Dy-d-Al-e-Ga & 0.0902 \\
Nd$_3$Dy & 139 & a-Dy-b-Nd-d-Nd & 0.0690 \\
Pd$_3$Pt & 139 & a-Pt-b-Pd-d-Pd & -0.0298 \\
LaMgSn & 139 & c-Mg-e-La-e-Sn & 0.0274 \\
Rb$_3$Mn$_2$F$_7$ & 139 & a-F-b-Rb-e-Rb-e-Mn-e-F-g-F & 0.0874 \\
AlAgAu$_2$ & 139 & a-Al-b-Ag-d-Au & 0.0912 \\
Ba$_3$Tl & 139 & a-Tl-b-Ba-d-Ba & 0.0933 \\
CsPrTe$_2$ & 166 & a-Cs-b-Pr-c-Te & 0.0977 \\
Sr$_2$Ca & 166 & a-Ca-c-Sr & 0.0771 \\
Pr$_2$AlRu$_3$ & 166 & a-Al-c-Pr-d-Ru & 0.0909 \\
Dy$_3$Ho & 194 & d-Ho-h-Dy & 0.0738 \\
Zr$_3$Pb & 194 & d-Pb-h-Zr & 0.0426 \\
PrAu$_3$ & 194 & d-Pr-h-Au & 0.0696 \\
Pd$_3$Pt & 194 & d-Pt-h-Pd & 0.0053 \\
TbZnGa & 194 & a-Tb-c-Ga-d-Zn & 0.0718 \\
BaSr(GaGe)$_2$ & 194 & a-Ba-b-Sr-f-Ga-f-Ge & 0.0855 \\
Ho$_2$Er & 194 & b-Er-f-Ho & 0.0790 \\
NdDyHg$_2$ & 225 & a-Nd-b-Dy-c-Hg & 0.0974 \\
MgAgPd$_2$ & 225 & a-Mg-b-Ag-c-Pd & 0.0540 \\
PrEuIn$_2$ & 225 & a-Eu-b-Pr-c-In & 0.0957 \\
TaNbRu$_2$ & 225 & a-Ta-b-Nb-c-Ru & 0.0823 \\
PdAu & 221 & a-Au-b-Pd & 0.0025 \\
Gd$_2$HgAu & 225 & a-Hg-b-Au-c-Gd & 0.0522 \\
\bottomrule
\label{table:samples}
\end{tabular}
\begin{tablenotes}
    \item[1] The fractional coordinates and lattice parameters are omitted for brevity. See details at \href{https://drive.google.com/file/d/1gOIkWkjSH_Ed0-wzPk8VgxLjJcidbrH3/view?usp=sharing}{https://drive.google.com/file/d/1gOIkWkjSH\_Ed0-wzPk8VgxLjJcidbrH3/view?usp=sharing}.
\end{tablenotes}
\end{threeparttable}
\end{table}






\begin{table}[t]
\begin{threeparttable}
\centering
\caption{Discovered crystalline materials by PyXtal initialization with $E_\mathrm{hull} < 0.1$ eV/atom which are not in the MP-20 dataset.}
\renewcommand{\arraystretch}{1.2}
\begin{tabular}{lclr}
\toprule
\textbf{Formula} & \textbf{Space group} & \textbf{Wyckoff-Atom sequence}~\tnote{1}  &  \textbf{$E_\mathrm{hull}$ (eV/atom)} \\
\midrule

Ba$_2$Si & 62 & b-Ba-c-Ba-c-Si & 0.0766 \\
GdIr & 62 & c-Gd-c-Ir & -0.2310 \\
Nd$_3$Dy & 139 & a-Dy-b-Nd-d-Nd & 0.0680 \\
Pd$_3$Pt & 139 & a-Pd-b-Pt-d-Pd & 0.0147 \\
PdPt & 221 & a-Pd-b-Pt & 0.0847 \\
NdDyHg$_2$ & 225 & a-Dy-b-Nd-c-Hg & 0.0973 \\
MgAgPd$_2$ & 225 & a-Mg-b-Ag-c-Pd & 0.0480 \\
HfRe & 221 & a-Re-b-Hf & 0.0926 \\
TaNbRu$_2$ & 225 & a-Nb-b-Ta-c-Ru & 0.0770 \\
PdAu & 221 & a-Au-b-Pd & 0.0056 \\
Gd$_2$HgAu & 225 & a-Hg-b-Au-c-Gd & 0.0476 \\
TbMgPd$_2$ & 225 & a-Mg-b-Tb-c-Pd & 0.0676 \\

\bottomrule
\label{table:pyxtal_samples}
\end{tabular}
\begin{tablenotes}
    \item[1] The fractional coordinates and lattice parameters are omitted for brevity. See details at \href{https://drive.google.com/file/d/18sdsp-6yRSBaNez1A0lr20Ru6H7OOzvJ/view?usp=drive_link}{https://drive.google.com/file/d/18sdsp-6yRSBaNez1A0lr20Ru6H7OOzvJ/view?usp=drive\_link}. 
\end{tablenotes}
\end{threeparttable}
\end{table}
    

\clearpage
\newpage

\section{Discovered crystal samples with element substitution}
\label{sec:mutation-app}
Tables~\ref{table:mutation_samples} and \ref{table:mutation_pmg_samples} list double perovskites crystals in the $Fm\bar{3}m$ (No. 225) space group with $E_\mathrm{hull} < 0.1$ eV/atom found by \texttt{CrystalFormer} and element substitution of pymatgen respectively. We found that \texttt{CrystalFormer} has found three times more  perovskites crystals compared to ionic substitution~\cite{hautier2011data}.
The DFT calculation setup is the same as in the previous section.


\begin{table}[h]
\begin{threeparttable}
\centering
\caption{Discovered double perovskites by likelihood guided structure mutation of \texttt{CrystalFormer} with $E_\mathrm{hull} < 0.1$ eV/atom which are not in the MP-20 dataset.}
\renewcommand{\arraystretch}{1.2}
\begin{tabular}{lclcc}
\toprule
\textbf{Formula} & \textbf{Space group} & \textbf{Wyckoff-Atom sequence}~\tnote{1}  &  \textbf{$E_\mathrm{hull}$ (eV/atom)} &  \textbf{Lattice constant ($\AA$)}  \\
\midrule

Sr$_2$TaNiO$_6$ & 225 & a-Ni-b-Ta-c-Sr-e-O & 0.0426 & 7.9762 \\
Ba$_2$TaNiO$_6$ & 225 & a-Ni-b-Ta-c-Ba-e-O & 0.0711 & 8.2129 \\
Ba$_2$YMnO$_6$ & 225 & a-Y-b-Mn-c-Ba-e-O & 0.0284 & 8.3575 \\
Eu$_2$CoMoO$_6$ & 225 & a-Co-b-Mo-c-Eu-e-O & 0.0720 & 7.9220 \\
Ba$_2$TbFeO$_6$ & 225 & a-Tb-b-Fe-c-Ba-e-O & 0.0343 & 8.3200 \\
Ba$_2$HfWO$_6$ & 225 & a-W-b-Hf-c-Ba-e-O & 0.0939 & 8.3568 \\
Eu$_2$CoNiO$_6$ & 225 & a-Ni-b-Co-c-Eu-e-O & 0.0646 & 7.6181 \\
Sr$_2$NbVO$_6$ & 225 & a-V-b-Nb-c-Sr-e-O & 0.0166 & 8.0375 \\
Sr$_2$NaWO$_6$ & 225 & a-Na-b-W-c-Sr-e-O & 0.0901 & 8.2959 \\ 

\bottomrule
\label{table:mutation_samples}
\end{tabular}
\begin{tablenotes}
    \item[1] The fractional coordinates are omitted for brevity.
\end{tablenotes}
\end{threeparttable}
\end{table}

\begin{table}[h]
\begin{threeparttable}
\centering
\caption{Discovered double perovskites using the element substitution of Hautier et al.~\cite{hautier2011data} with $E_\mathrm{hull} < 0.1$ eV/atom which are not in the MP-20 dataset.}
\renewcommand{\arraystretch}{1.2}
\begin{tabular}{lclcc}
\toprule
\textbf{Formula} & \textbf{Space group} & \textbf{Wyckoff-Atom sequence}~\tnote{1}  &  \textbf{$E_\mathrm{hull}$ (eV/atom)} &  \textbf{Lattice constant ($\AA$)}  \\
\midrule

Ba$_2$FeBiO$_6$ & 225 & a-Bi-b-Fe-c-Ba-e-O & -0.0030 & 8.3808 \\
Ba$_2$NpCoO$_6$ & 225 & a-Np-b-Co-c-Ba-e-O & 0.0241 & 8.3983 \\
Ba$_2$BiMoO$_6$ & 225 & a-Bi-b-Mo-c-Ba-e-O & 0.0706 & 8.6322 \\

\bottomrule
\label{table:mutation_pmg_samples}
\end{tabular}
\begin{tablenotes}
    \item[1] The fractional coordinates are omitted for brevity.
\end{tablenotes}
\end{threeparttable}
\end{table}

\clearpage
\newpage
\section{Details of plug-and-play materials design}
\label{sec:plug-and-play-app}

We use two pre-trained MEGNet models~\cite{chen2019graph,chen2021learning} for band gap and formation energy prediction, which is implemented in the MatGL~\cite{Ko_Materials_Graph_Library_2021}
repository. It is important to note that we do not train the models from scratch or fine-tune them on the MP20 dataset. Instead, we utilize 
the pre-trained models provided in the repository. This approach closely aligns with actual practices in materials science, where numerous 
high-quality machine learning property predictors are readily accessible.

For the regression task, the mean absolute error corresponds to a predictive model $f(\boldsymbol{\mathcal{C}})$ characterized by the Laplace distribution $ p(y|\boldsymbol{\mathcal{C}} ) \propto \exp(-\alpha |y - f(\boldsymbol{\mathcal{C}})| ) $, where $\alpha$ is a positive scale parameter. According to Bayes' rule, the posterior probability for crystals given the property $y$ is:
\begin{equation}
    p(\boldsymbol{\mathcal{C}}|g, y) \propto p(y|\boldsymbol{\mathcal{C}})p(\boldsymbol{\mathcal{C}}|g) \propto e^{-\alpha|y - f(\boldsymbol{\mathcal{C}})| } p(\boldsymbol{\mathcal{C}}|g).
\end{equation}
The crystal log-likelihood with a given property condition is then given by:
\begin{equation}
\ln p(\boldsymbol{\mathcal{C}}|g, y) \propto \ln p(\boldsymbol{\mathcal{C}}|g) - \alpha |y - f(\boldsymbol{\mathcal{C}})|. 
\end{equation}
It is instructive to see that property-guided samples are balanced between the \texttt{CrystalFormer} prior and the property prediction model likelihood. Here $\alpha $ plays the role of guidance strength. For multi-property prediction, the likelihood function can be extended to:
\begin{equation}
\ln p(\boldsymbol{\mathcal{C}}|g, y_{1}, \ldots, y_{n}) \propto  \ln p(\boldsymbol{\mathcal{C}}|g) - \sum_{i=1}^{n} \alpha_{i} |y_{i} - f_{i}(\boldsymbol{\mathcal{C}})|.
\end{equation}

Note that MCMC sampling via the Metropolis algorithm does not involve the normalization factor of the conditional distribution.
In our experiments, we set $\alpha_{1} = 3$ for band gap prediction and $\alpha_{2} = 10$ for formation energy prediction, corresponding to the inverses of the MAE values obtained from the regression model.

Table \ref{table:plug_and_play_samples} shows a list of novel samples close to the energy hull with formation energy and band predicted by the MEGNet~\cite{chen2019graph,chen2021learning}  and DFT calculations.

\begin{table}[t]
    \begin{threeparttable}
    \centering
    \caption{Samples discovered by plug-and-play manner with $E_\mathrm{hull} < 0.1$ eV/atom which are not in the MP-20 dataset. The energy above the convex hull is computed via DFT while the formation energy and band gap are predicted both by DFT and machine learning model.}
    \renewcommand{\arraystretch}{1.2}
    \begin{tabular}{lclccccc}
    \toprule
    
    \multirow{1}{*}{\textbf{Formula}} & \multirow{1}{*}{\textbf{Space group}} & 
    \multirow{1}{*}{\textbf{Wyckoff-Atom sequence}}  & \multirow{1}{*}{\textbf{$E_\mathrm{hull}$ (eV/atom)}} &
     \multicolumn{2}{c}{\textbf{Formation energy (eV/atom)}} & \multicolumn{2}{c}{\textbf{Band gap (eV)}}  \\
    & &  &  & DFT & MEGNet & DFT & MEGNet  \\
    
    \midrule
    
    LiMoF$_6$ & 225 & a-Mo-b-Li-e-F & -0.4816 & -3.2893 & -2.8811 & 0.0039 & 2.3289 \\
    Sr$_2$ZrWO$_6$ & 225 & a-Zr-b-W-c-Sr-e-O & -0.4642 & -3.6203 & -3.0082 & 0.0030 & 1.6535 \\
    Sr$_2$HfWO$_6$ & 225 & a-Hf-b-W-c-Sr-e-O & -0.4222 & -3.6537 & -3.1408 & 0.0126 & 1.9169 \\
    K$_2$NaWF$_6$ & 225 & a-W-b-Na-c-K-e-F & -0.2964 & -3.1146 & -2.8592 & 1.2693 & 1.8388 \\
    Cs$_2$RbMoF$_6$ & 225 & a-Mo-b-Rb-c-Cs-e-F & -0.2905 & -3.1418 & -2.9865 & 2.2306 & 1.8929 \\
    KCrF$_6$ & 225 & a-K-b-Cr-e-F & -0.2802 & -2.9034 & -2.8221 & 0.0006 & 2.3546 \\
    Ba$_2$YMnO$_6$ & 225 & a-Y-b-Mn-c-Ba-e-O & -0.2439 & -3.1016 & -2.7970 & 0.0028 & 2.0245 \\
    Ba$_2$ZrMnO$_6$ & 225 & a-Zr-b-Mn-c-Ba-e-O & -0.2082 & -3.2689 & -3.0276 & 0.9285 & 1.8680 \\
    Sr$_2$HfMnO$_6$ & 225 & a-Hf-b-Mn-c-Sr-e-O & -0.1649 & -3.3213 & -3.0912 & 0.8717 & 1.8699 \\
    Ba$_2$LaVO$_6$ & 225 & a-V-b-La-c-Ba-e-O & -0.1624 & -3.3785 & -3.0244 & 1.3322 & 1.5306 \\
    Cs$_2$RbCrF$_6$ & 225 & a-Rb-b-Cr-c-Cs-e-F & -0.1562 & -3.1799 & -2.9984 & 2.2985 & 3.2548 \\
    Cs$_2$RbVF$_6$ & 225 & a-V-b-Rb-c-Cs-e-F & -0.1356 & -3.2064 & -3.1524 & 0.0014 & 1.7538 \\
    Cs$_2$NaVF$_6$ & 225 & a-Na-b-V-c-Cs-e-F & -0.1104 & -3.2037 & -3.1418 & 0.0004 & 2.0226 \\
    Cs$_2$MnTlF$_6$ & 225 & a-Mn-b-Tl-c-Cs-e-F & -0.0666 & -2.8449 & -2.8082 & 0.0010 & 2.2036 \\
    K$_2$MnCoF$_6$ & 225 & a-Mn-b-Co-c-K-e-F & -0.0513 & -2.8679 & -2.7951 & 0.0038 & 2.1415 \\
    K$_2$RbVF$_6$ & 225 & a-V-b-Rb-c-K-e-F & -0.0159 & -3.1498 & -3.0416 & 0.0010 & 1.5841 \\
    HoMgPt$_2$ & 225 & a-Mg-b-Ho-c-Pt & 0.0212 & -0.9985 & -1.0616 & 0.0215 & -0.0074 \\
    TmPt & 221 & a-Pt-b-Tm & 0.0563 & -1.2660 & -1.2784 & 0.1493 & -0.0091 \\
    K$_2$SrCoF$_6$ & 225 & a-Sr-b-Co-c-K-e-F & 0.0706 & -3.0990 & -3.1450 & 0.0018 & 1.1738 \\
    HoTmIr$_2$ & 225 & a-Tm-b-Ho-c-Ir & 0.0712 & -0.8510 & -0.8649 & 0.0125 & -0.0074 \\
    Li$_2$VF$_6$ & 225 & a-V-c-Li-e-F & 0.0729 & -3.0782 & -2.9730 & 0.0015 & 1.8764 \\
    Lu$_2$IrPt & 225 & a-Pt-b-Ir-c-Lu & 0.0788 & -1.0960 & -1.1430 & 0.0047 & -0.0076 \\
    Cs$_2$LiCrF$_6$ & 225 & a-Cr-b-Li-c-Cs-e-F & 0.0843 & -3.0344 & -2.9774 & 1.3975 & 1.9886 \\
    ErLuPt$_2$ & 225 & a-Lu-b-Er-c-Pt & 0.0850 & -1.2383 & -1.2755 & 0.0066 & -0.0098 \\
    HoLuPt$_2$ & 225 & a-Lu-b-Ho-c-Pt & 0.0944 & -1.2223 & -1.2622 & 0.0327 & -0.0102 \\
    ErPt & 221 & a-Pt-b-Er & 0.0951 & -1.2140 & -1.2607 & 0.0691 & -0.0089 \\

\bottomrule
\label{table:plug_and_play_samples}
\end{tabular}
\end{threeparttable}
\end{table}

\end{document}